\documentclass[aps,superscriptaddress,reprint,nofootinbib]{revtex4-1}

\usepackage{graphicx}
\usepackage{amsmath}
\usepackage{amssymb}
\usepackage{amsfonts}
\usepackage{filecontents}
\usepackage{color}
\usepackage[normalem]{ulem} 
\usepackage{soul}
\usepackage{epstopdf}
\usepackage{nameref}
\usepackage{hyperref}
\hypersetup{hidelinks}
\usepackage{fancyhdr}
\usepackage{float}
\usepackage{gensymb}
\usepackage{physics}
\usepackage{upgreek}
\newcounter{fig}
\usepackage{lipsum}
\usepackage{natbib}
\usepackage{bibunits} 
\usepackage[usenames,dvipsnames,svgnames,table]{xcolor}
\usepackage{enumitem}
\usepackage{multirow}

\DeclareRobustCommand{\SkipTocEntry}[5]{}

\defaultbibliography{Auger_paper_reference.bib}
\renewcommand{\bibliography}[1]{} 
\begin{document}
\begin{bibunit}[naturemag]

\title{Coherent control of a high-orbital hole in a semiconductor quantum dot}

\author{Jun-Yong Yan}
\author{Chen Chen}
\author{Xiao-Dong Zhang}
\author{Yu-Tong Wang}
\affiliation{Interdisciplinary Center for Quantum Information, State Key Laboratory of Extreme Photonics and Instrumentation, College of Information Science and Electronic Engineering, Zhejiang University, Hangzhou, China}

\author{Hans-Georg Babin}

\author{\\Andreas D. Wieck}
\author{Arne Ludwig}
\affiliation{Lehrstuhl für Angewandte Festkörperphysik, Ruhr-Universität Bochum, Bochum, Germany}
\author{Yun Meng}
\author{Xiaolong Hu}
\affiliation{School of Precision Instrument and
Optoelectronic Engineering, Tianjin University, Tianjin, China}
\affiliation{Key Laboratory of Optoelectronic
Information Science and Technology, Ministry of Education, Tianjin, China}

\author{Huali Duan}
\author{Wenchao Chen}
\affiliation{Interdisciplinary Center for Quantum Information, State Key Laboratory of Extreme Photonics and Instrumentation, College of Information Science and Electronic Engineering, Zhejiang University, Hangzhou, China}
\affiliation{ZJU-UIUC Institute, International Campus, Zhejiang University, Haining, China}

\author{\\Wei Fang}
\affiliation{College of Optical Science and
Engineering, Zhejiang University, Hangzhou, China}

\author{Moritz Cygorek}
\affiliation{SUPA, Institute of Photonics and Quantum Sciences, Heriot-Watt University, Edinburgh EH14 4AS, UK}
 
\author{Xing Lin}
\affiliation{Interdisciplinary Center for Quantum Information, State Key Laboratory of Extreme Photonics and Instrumentation, College of Information Science and Electronic Engineering, Zhejiang University, Hangzhou, China}
\author{Da-Wei Wang}
\affiliation{Zhejiang Province Key Laboratory of Quantum Technology and Device, School of Physics, Zhejiang University, Hangzhou, China}
\author{Chao-Yuan Jin}
\affiliation{Interdisciplinary Center for Quantum Information, State Key Laboratory of Extreme Photonics and Instrumentation, College of Information Science and Electronic Engineering, Zhejiang University, Hangzhou, China}
\affiliation{International Joint Innovation Center, Zhejiang University, Haining, China}
\affiliation{Center for Information Technology Application Innovation, Shaoxing Institute, Zhejiang University, Shaoxing, China}

\author{Feng Liu}
\email[Email to: ]{feng\_liu@zju.edu.cn}
\affiliation{Interdisciplinary Center for Quantum Information, State Key Laboratory of Extreme Photonics and Instrumentation, College of Information Science and Electronic Engineering, Zhejiang University, Hangzhou, China}
\affiliation{International Joint Innovation Center, Zhejiang University, Haining, China}


\begin{abstract}
\textbf{Coherently driven semiconductor quantum dots are one of the most promising platforms for non-classical light sources and quantum logic gates which form the foundation of photonic quantum technologies. However, to date, coherent manipulation of single charge carriers in quantum dots is limited mainly to their lowest orbital states. Ultrafast coherent control of high-orbital states is obstructed by the demand for tunable terahertz pulses. To break this constraint, we demonstrate an all-optical method to control high-orbital states of a hole via stimulated Auger process. The coherent nature of the Auger process is proved by Rabi oscillation and Ramsey interference. Harnessing this coherence further enables the investigation of single-hole relaxation mechanism. A hole relaxation time of 161~ps is observed and attributed to the phonon bottleneck effect. Our work opens new possibilities for understanding the fundamental properties of high-orbital states in quantum emitters and developing new types of orbital-based quantum photonic devices.} 
\end{abstract}

\maketitle

Coherent control of a quantum emitter lies at the heart of quantum optics. A variety of coherent quantum optical processes, such as Rabi oscillation~\cite{Zrenner2002a}, stimulated Raman transition~\cite{Press2008}, electromagnetically induced transparency~\cite{Fleischhauer2005} and photon blockade~\cite{Faraon2008},
have been demonstrated with quantum emitters. Coherently driven solid-state quantum emitters, e.g. semiconductor quantum dots (QDs), are particularly attractive for realizing compact and scalable quantum photonic devices, including on-demand quantum light sources~\cite{Michler2000,Ding2016c,Tomm2021b,Liu2018f,Huber2017c,Liu2019e}, single-photon all-optical switches~\cite{Sun2018,Jeannic2022}, and photonic quantum logic gates~\cite{Li2003}. These devices form the basis of photonic quantum computing, communication and sensing~\cite{Pelucchi2021}.

\begin{figure*}
\refstepcounter{fig}
    \includegraphics[width=0.95\textwidth]{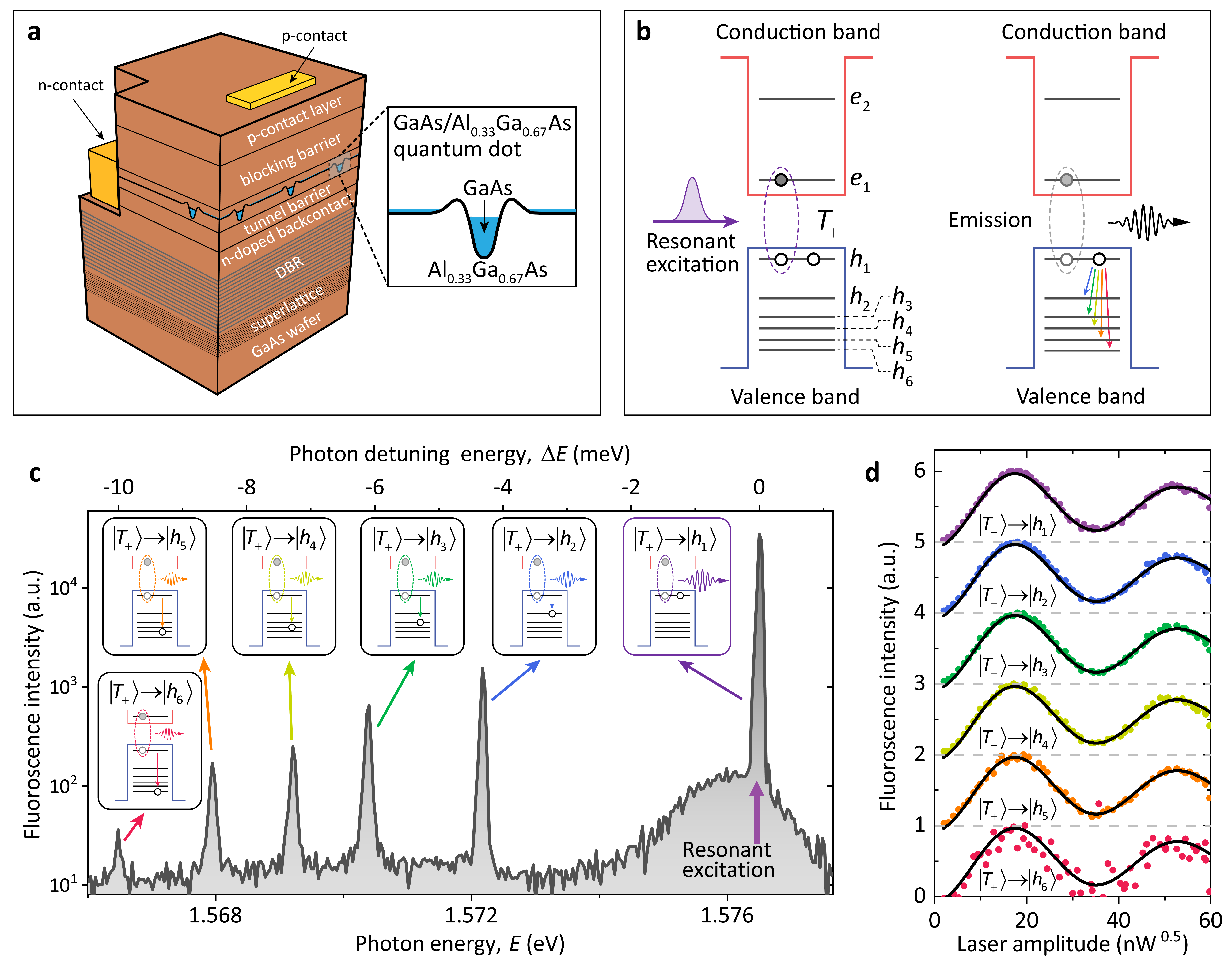} 
    \caption{ \textbf{Radiative Auger emission from a positively charged QD.} \textbf{a}, The heterostructure of the sample. A low-density layer of droplet epitaxy GaAs QDs is embedded in an n-i-p diode. DBR: distributed Bragg reflector. \textbf{b}, Energy-level diagram of the QD. Left: resonant excitation. A pulse resonant with the $\left|h_1\right>\leftrightarrow\left|\textit{T}_\text{+}\right>$ transition creates a positive trion $\textit{T}_\text{+}$ composed of an electron and two holes. Right: radiative recombination channels. After the recombination of an electron-hole pair, the QD emits a zero-detuning photon or a red-shifted Auger photon accompanied by exciting the residual hole to a high-orbital level. \textbf{c}, Fluorescence spectrum under resonant excitation (thick purple arrow). The highest peak at zero detuning originates from the fundamental transition ($\left|\textit{T}_\text{+}\right>\rightarrow\left|h_1\right>$). A series of satellite peaks at the lower energy side correspond to the radiative Auger transitions to high-orbital hole states. Insets: schematic description of emission processes. \textbf{d}, Normalized fluorescence intensity of each emission line in \textbf{c} as a function of the laser amplitude resonant with the ${\left|h_1\right>}\leftrightarrow{\left|\textit{T}_\text{+}\right>}$ transition, exhibiting a series of Rabi oscillations with the same period. Black lines: global fitting with a damped sinusoidal function. a.u., arbitrary units.}
\label{Observation}
\end{figure*}

Quantum dots, also called artificial atoms, are semiconductor nanostructures with discrete energy levels of electrons and holes (see Figs.~\ref{Observation}a, b)~\cite{Bayer2002c}. Up to now, coherent manipulation of single charge carriers in QDs is limited mainly to their lowest orbital states ($\left|e_1\right>$ and $\left|h_1\right>$). Coherent control of high-orbital states is highly desirable because this capability would enable a wide range of fundamental research and applications. It would provides an excellent platform to investigate single-carrier relaxation mechanisms~\cite{Zibik2009,Lobl2020}, which is crucial for improving the brightness of QD-based lasers and quantum light sources. Such orbital states could also be used for building new types of optically active solid-state quantum bits (qubits) for quantum information processing. Furthermore, by controlling the spatial distribution of charge carrier wavefunctions~\cite{Qian2019}, one could realize a strongly coupled quantum emitter-optical cavity system ~\cite{Volz2012}, a key component for quantum networks~\cite{Kimble2008}. 

However, since the spacings between electron/hole levels in QDs are in the terahertz (THz) regime, coherent control of high-orbital states requires tunable transform-limited pico- or femotosecond THz pulses typically generated from free electron lasers ~\cite{Zibik2009,Litvinenko2015} which are inaccessible for table-top applications. Despite considerable efforts to develop solid-state THz lasers~\cite{Qin2009,Taschler2021}, a laser simultaneously meeting all the above requirements is still commercially unavailable. Additionally, the long wavelength of THz radiation makes it difficult to address a single nanoscale quantum emitter. A fully optical coherent control scheme can release this constraint and unlock the potential of high-orbital states.

Recently, the radiative Auger process, which was initially discovered in atoms~\cite{Aberg1969} and later in solid-state emitters such as quantum wells~\cite{Nash1993}, ensembles of quantum dots~\cite{Paskov2000a}, and colloidal nanoplatelets~\cite{Antolinez2019}, emerged as a possible tool to generate high-orbital charge carriers. In 2020, Löbl et al. reported the observation of the radiative Auger process in a single negatively charged QD~\cite{Lobl2020}. In this process, an electron and a hole recombine, emit a photon and simultaneously excite a residual electron to a high-orbital state. Driving the radiative Auger transitions with a continuous-wave (CW) laser led to the observation of electromagnetically induced transparency (EIT)~\cite{Spinnler2021,Fleischhauer2005}. Although these results indicate a great potential of the radiative Auger process in preparing high-orbital carriers, deterministic generation and coherent control of a high-orbital carrier are yet to be demonstrated.

In this article, we demonstrate ultrafast coherent control of a high-orbital hole in a QD with near-unity fidelity ($95.9\%$) via \textit{stimulated} Auger process. This scheme provides an optical approach to manipulating orbital states with THz spacings, breaking the bottleneck caused by the hardly accessible tunable pulsed THz lasers. The barely explored coherent nature of the Auger process is proved by Rabi oscillation and Ramsey interference measurements. The coherence time is limited mainly by the hole relaxation time, indicating the great potential of hole orbital qubits for achieving long coherence times without the need for strong magnetic field. Different from driving the radiative Auger transition with a CW laser in a negatively charged QD~\cite{Lobl2020}, resonant picosecond pulsed excitation allows \textit{deterministic} generation and ultrafast \textit{coherent control} of a high-orbital hole in a positively charged QD. With this unique capability, we study the single-hole relaxation dynamics and observe the phonon bottleneck effect. Our work paves the way towards orbital-based solid-state quantum photonic devices.

\addtocontents{toc}{\SkipTocEntry}

\section*{Radiative Auger transitions in a positively charged quantum dot}

In the radiative Auger process, the recombination of an electron-hole pair is accompanied by emitting a photon and by exciting a residual charge carrier to a higher orbital state. In order to observe this process, it is necessary to prepare a charged QD first. We grow droplet epitaxy GaAs QDs hosted in an n-i-p diode as shown in Fig.~\ref{Observation}a. By properly choosing the bias, the QD can be optically charged with a single hole via fast electron tunneling (see details in Supplementary Section~I)~\cite{Tomm2021b}. 
A positive trion ($\textit{T}_\text{+}$) containing an electron and two holes is created by exciting the QD with a 6~ps pulse resonant to the $\left|h_1\right>\leftrightarrow\left|\textit{T}_\text{+}\right>$ transition (fundamental transition) as shown in Fig.~\ref{Observation}b. The trion then recombines radiatively and leaves a hole in the QD. The residual hole mostly stays in the lowest orbital $h_1$. When radiative Auger transitions occur in the recombination, a red-shifted photon is emitted and a hole is excited to a higher orbital, e.g. $h_2$, $h_3$, etc.

It should be noted that, while the conduction band states in the III-V semiconductor QDs are usually well approximated by harmonic oscillator eigenstates, this is often not the case for hole states because of the proximity of the heavy-hole and the light-hole bands. Energy eigenstates in the valence band are therefore easily mixed by the effects of the confinement potential, spin-orbital coupling, and strain~\cite{Cygorek2020,Zielinski2010,Holtkemper2018}. Thus, the orbital angular momentum is not a good quantum number and we choose to label the confined hole states in terms of their energetic order $h_1$, $h_2$, $h_3$, etc.~\cite{Reindl2019b} instead of harmonic oscillator label $s$, $p$, $d$, etc. The corresponding symmetry breaking is responsible not only for the more irregular level spacing, but also for the finite dipole of radiative Auger transitions~\cite{Gawarecki2022}.

The radiative processes mentioned above can be observed in the fluorescence spectra of the resonantly excited QD (see Fig.~\ref{Observation}c). The strongest peak at 1.5765~eV corresponds to the emission from the fundamental transition (see the purple inset). The weak sideband extended by $\sim-3$~meV originates from the phonon-assisted emission~\cite{Iles-Smith2017c,Roy2011,Reigue2017a}. Five satellite peaks at the lower energy side of the fundamental line correspond to photons generated via the radiative Auger process with the residual hole arriving at different orbitals (see the insets).  In the rest of the article, we refer to these satellite peaks as Auger peaks. Since radiative Auger transitions in QDs become optically allowed mainly due to slight structural asymmetry~\cite{Gawarecki2022}, the Auger peaks are 1-3 orders of magnitude weaker than the fundamental line. Their detunings (a few THz) relative to the fundamental line directly give the energy level spacings of hole orbital states. We note here that the radiative Auger process occurring via an excited real state is fundamentally different from a Raman process~\cite{Press2008,DeGreve2011a} which occurs via a virtual state.

To verify the origin of the Auger peaks, we sweep the resonant laser power and measure the intensity of all emission lines. Figure~\ref{Observation}d presents the oscillating fluorescence intensity of each emission line, resulting from the well-known Rabi rotation between $\left|\textit{T}_\text{+}\right>$ and $\left|h_1\right>$. The Rabi oscillations with the same period confirm that all emission lines originate not only from the same QD but also from the same initial state $\left|\textit{T}_\text{+}\right>$. This conclusion is cross-checked by the magneto-optical spectra showing anti-crossings of Auger peaks in high magnetic field (see Supplementary Fig.~5).

\begin{figure*}
\refstepcounter{fig}
    \includegraphics[width=0.95\linewidth]{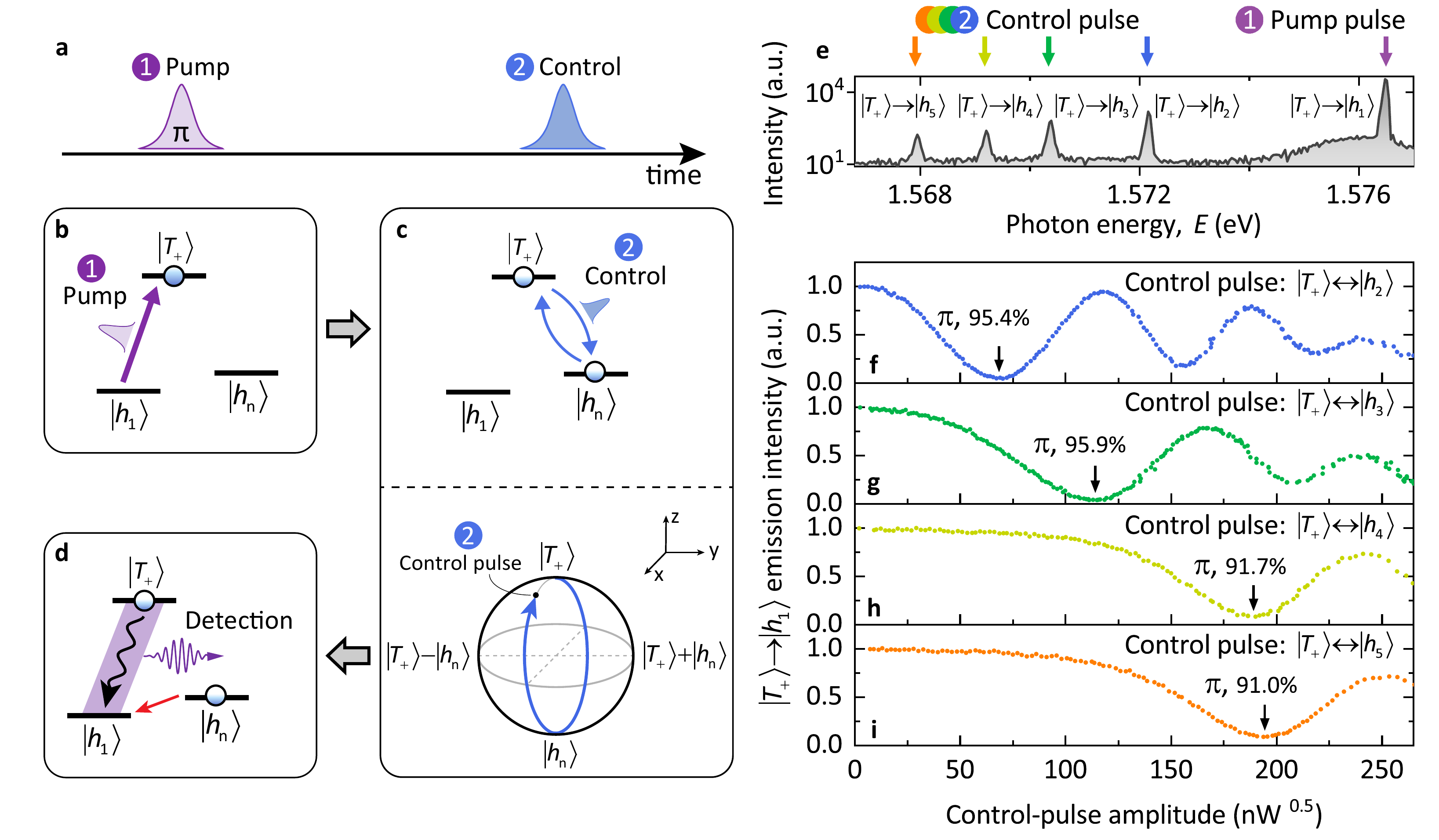}
    \caption{\textbf{Rabi oscillation of a high-orbital hole} \textbf{a}, The timing diagram of the pulse sequence used in this experiment. \textbf{b-d},  Principle of the measurement: \textbf{b}, Preparation of trion. Resonant excitation of the $\left|h_1\right>\leftrightarrow\left|\textit{T}_\text{+}\right>$ transition with a $\pi$ pump pulse creates a positive trion. $\left|h_n\right>$: high-orbital states, n=2, 3, 4, 5. \textbf{c}, Coherent control of radiative Auger transitions. The high-orbital hole states and trion are coupled via coherent Auger transitions driven by the control pulse. This process can be visualized as rotating the Bloch vector around the y-axis. \textbf{d}, Detection. The trion population is measured by detecting the $\left|\textit{T}_\text{+}\right>\rightarrow\left|h_1\right>$ emission. \textbf{e}, The excitation scheme illustrated in the fluorescence spectrum. Violet arrow: pump pulse. Blue, green, yellow, and orange arrows: the subsequent control pulse resonant with each Auger transition. \textbf{f-i}, Normalized $\left|T_+\right>\rightarrow\left|h_1\right>$ emission intensity, for control pulses resonant with $\left|T_+\right>\leftrightarrow\left|h_2\right>$ (\textbf{f}), $\left|h_3\right>$ (\textbf{g}), $\left|h_4\right>$ (\textbf{h}) and $\left|h_5\right>$ (\textbf{i}) transitions, as a function of the amplitude of the control pulse indicated in \textbf{e}. The $\left|\textit{T}_\text{+}\right>\rightarrow\left|h_1\right>$ emission reaches its ﬁrst dip at control-pulse area $\Theta=\pi$ (black arrows), indicating the deterministic preparation of high-orbital hole states with near-unity fidelity.}
    \label{Rabi}
\end{figure*}

\addtocontents{toc}{\SkipTocEntry}
\section*{Stimulated Auger transitions}

In the previous section, we present the evidence of the radiative Auger process which provides an optical channel to access orbital states. However, the spontaneous radiative Auger process occurs rarely compared with the fundamental transition (see Fig.~\ref{Observation}c), hindering its application. We overcome this drawback and bring the control of high-orbital hole states to a new level using \textit{stimulated} Auger transitions. This control is achieved by coherently driving the Auger process with a picosecond optical pulse resonant to transitions between the trion and high-orbital hole states. Under this condition, the originally weak Auger process could occur within the duration of the ultrafast optical pulse with high fidelity.

Figures~\ref{Rabi}a-d illustrate our coherent control scheme. For clarity, trion/hole states are now represented as single energy levels. The QD is initially filled with a hole in the $h_1$ orbital. A resonant pump pulse with area $\Theta=\pi$ creates a positive trion $\textit{T}_\text{+}$. Subsequently, a control pulse coherently drives the Auger transition between $\left|\textit{T}_\text{+}\right>$ and high-orbital states (Fig.~\ref{Rabi}c), leading to a transfer of $\left|\textit{T}_\text{+}\right>$ population to high-orbital hole states. Finally, the remaining $\left|\textit{T}_\text{+}\right>$ population is measured through the $\left|\textit{T}_\text{+}\right>\rightarrow\left|h_1\right>$ emission intensity (Fig.~\ref{Rabi}d). The successful coherent control of a high-orbital hole is signaled by the quenching of the $\left|\textit{T}_\text{+}\right>\rightarrow\left|h_1\right>$ emission. The energies of the pump and control pulses are indicated by colored arrows in the fluorescence spectrum shown in Fig.~\ref{Rabi}e. The delay time between two pulses is 18 ps. The delay is small compared with the radiative lifetime of $\left|\textit{T}_\text{+}\right>$ (400(2)~ps, see Supplementary Fig.~7), and large enough to avoid pulse overlapping (see Fig.~\ref{Rabi}a), which is crucial for excluding the EIT effect~\cite{Fleischhauer2005}. 

Figures~\ref{Rabi}f-i show the measured intensities of the $\left|\textit{T}_\text{+}\right>\rightarrow\left|h_1\right>$ emission as a function of the control-pulse amplitude. Rabi oscillations between $\left|\textit{T}_\text{+}\right>$ and up to four high-orbital states are observed, proving the coherent nature of the Auger process. The $\left|\textit{T}_\text{+}\right>\rightarrow\left|h_1\right>$ emission intensity reaches its first dip at control-pulse area $\Theta=\pi$, heralding a stimulated Auger emission and deterministic generation of a high-orbital hole with a maximum fidelity of $95.9\%$. We note that the Rabi oscillations are aperiodic due to the alternating current (AC) Stark shift induced by the control laser, which can be reproduced by our numerical simulation (see Supplementary Section~IV).

\begin{figure*}
\refstepcounter{fig}
	\includegraphics[width=0.9\linewidth]{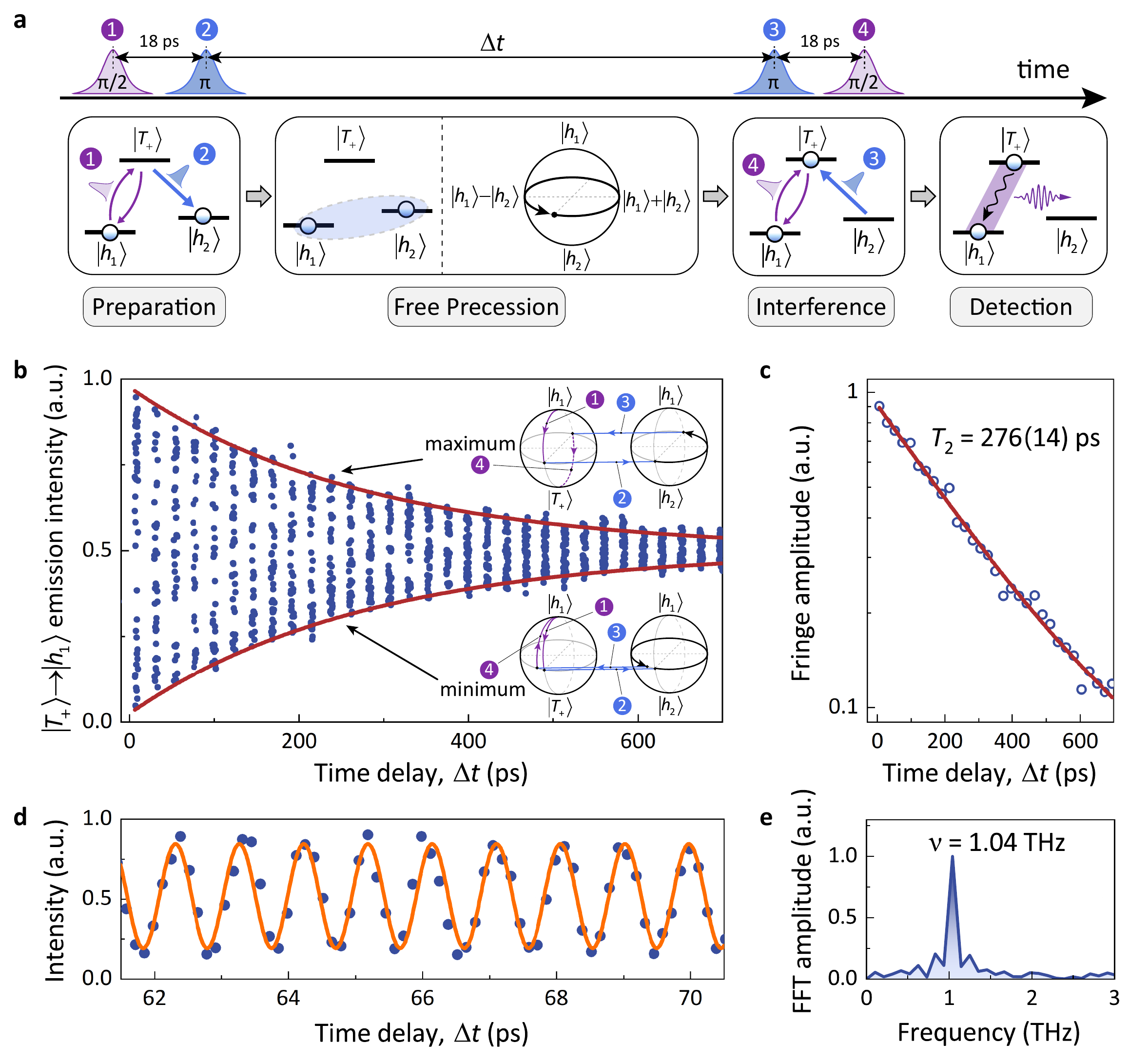}
	\caption{\textbf{Ramsey interference.} \textbf{a}, The measurement procedure. The first $\pi/2$ pulse and $\pi$ pulse (violet pulse 1 and blue pulse 2) separated by 18 ps drive the QD to a superposition state: $\left|\psi\right>=\frac{1}{\sqrt{2}}(\left|h_1\right>+i\left|h_2\right>)$. The evolution of the quantum state is probed by applying a delay-varied $\pi$ pulse and $\pi/2$ pulse (blue pulse 3 and violet pulse 4) and then detecting the $\left|\textit{T}_\text{+}\right>\rightarrow\left|h_1\right>$ emission. \textbf{b}, Ramsey fringes. The $\left|\textit{T}_\text{+}\right>\rightarrow\left|h_1\right>$ emission intensity as a function of $\Delta\textit{t}$. Top (bottom) inset: the evolution of the Bloch vector for the constructive (destructive) interference. Red brown lines: simulated interference envelope from a model based on the Bloch equation (Supplementary Section~VII). \textbf{c}, Decay of the fringe amplitude as a function of coarse delay, yielding a coherence time $\textit{T}_\text{2}=276(14)~\text{ps}$. Red brown line: fitting with a single exponential function. \textbf{d}, Ramsey fringes on a smaller time scale. Orange curve: fitting with a sinusoidal function. \textbf{e}, Fourier transform of the data in \textbf{d}, showing an oscillation frequency $\nu$ = 1.04 THz. FFT: fast Fourier transform.}
	\label{Ramsey}
\end{figure*}

\addtocontents{toc}{\SkipTocEntry}
\section*{Coherent control of a high-orbital hole}

Based on stimulated Auger transitions between trion and hole orbital states, we can realize coherent control, especially a superposition, between hole orbital states with THz energy differences. To this end, we measure the Ramsey interference~\cite{Press2008,DeGreve2011a,Greilich2011,Litvinenko2015,Godden2012,Greve2013} between $\left|h_1\right>\textrm{ and }\left|h_2\right>$, and evaluate their coherence time. Figure~\ref{Ramsey}a shows the principle of the measurement involving two pairs of two-color pulses. The first $\pi/2$ pulse (labeled No.~1) creates a superposition state, $\left|\psi_0\right>=\frac{1}{\sqrt{2}}(\left|h_1\right>+i\left|\textit{T}_\text{+}\right>)$. A subsequent $\pi$ pulse (labeled No.~2) transfers the whole population of $\left|\textit{T}_\text{+}\right>$ to $\left|h_2\right>$, leading to a superposition state: $\left|\psi_1\right>=\frac{1}{\sqrt{2}}(\left|h_1\right>+i\left|h_2\right>)$. Then the superposition state undergoes a free procession during $\Delta t$ with a frequency $\nu$ equals to the frequency difference between $\left|h_1\right>$ and $\left|h_2\right>$.
The third and fourth pulses (labeled No.~3 and~4) drive the QD towards $\left|\textit{T}_\text{+}\right>$ or $\left|h_1\right>$ depending on the phase accumulation (see Fig.~\ref{Ramsey}b insets). Finally, the $\left|\textit{T}_\text{+}\right>$ population is evaluated by measuring the $\left|\textit{T}_\text{+}\right>\rightarrow\left|\textit{h}_\text{1}\right>$ emission intensity.

\begin{figure*}
\refstepcounter{fig}
	\includegraphics[width=0.9\linewidth]{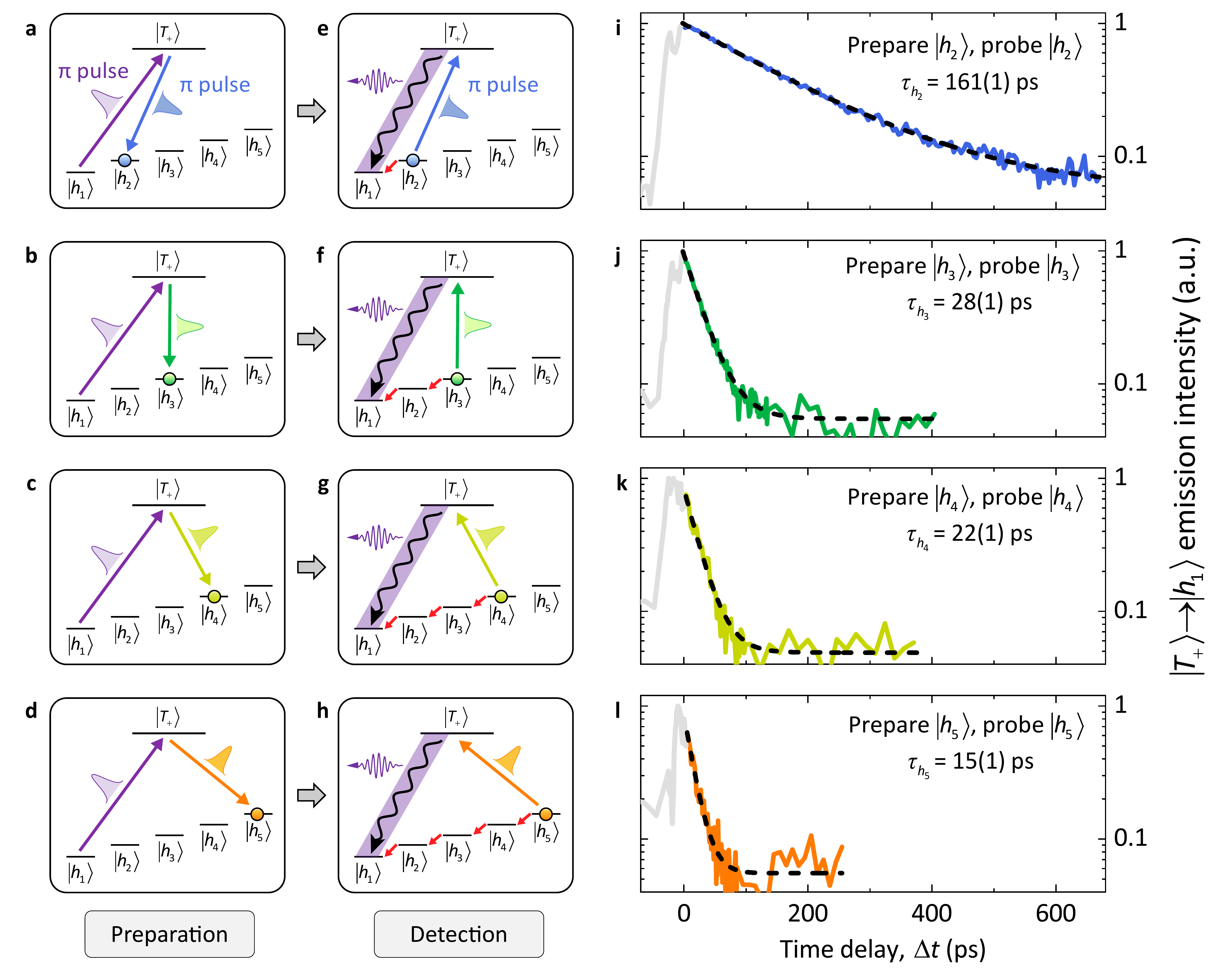}
	\caption{\textbf{Direct measurement of single-hole relaxation dynamics.} \textbf{a-h}, The measurement principle. \textbf{a-d}, A high-orbital hole is prepared deterministically by two pulses with area $\Theta=\pi$ via stimulating $\left|T_+\right>\rightarrow\left|h_2\right>$ (\textbf{a}), $\left|h_3\right>$ (\textbf{b}), $\left|h_4\right>$ (\textbf{c}) and $\left|h_5\right>$ (\textbf{d}) transitions. \textbf{e-h}, Then the population of the high-orbital hole state decays via phonon-assisted relaxation processes (red arrows). Finally, the remaining population of $\left|h_2\right>$ (\textbf{e}), $\left|h_3\right>$ (\textbf{f}), $\left|h_4\right>$ (\textbf{g}) and $\left|h_5\right>$ (\textbf{h}) is transferred to $\left|T_+\right>$ by the third $\pi$ pulse and measured by detecting the $\left|T_+\right>$ emission. \textbf{i-l}, The population of $\left|h_2\right>$ (\textbf{i}), $\left|h_3\right>$ (\textbf{j}), $\left|h_4\right>$ (\textbf{k}) and $\left|h_5\right>$ (\textbf{l}) as a function of the time delay $\Delta\textit{t}$ between the second and third $\pi$ pulses. The black dotted line shows the fitting with a single exponential function.} 
	\label{decay}
\end{figure*}

Interference fringes observed by varying the delay time between two pairs of two-color pulses with coarse and fine steps (see Figs.~\ref{Ramsey}b and d) reconfirm that the Auger transition is a coherent process. A coherence time of $\textit{T}_\text{2}=276(14)~\text{ps}$ is extracted by fitting the amplitude of the Ramsey fringes as a function of the coarse delay (Fig.~\ref{Ramsey}c). There are two effects responsible for the dephasing, namely population decay of $\left|h_2\right>$ with a lifetime $T_1 = \tau_{h_2}$ (161~ps, see Fig.~\ref{decay}i) and pure dephasing time $\textit{T}_{2}^{*}$. According to $1/T_2=1/T_2^*+1/(2T_1)$, $T_2^*=1.93$ ns is obtained, suggesting that the coherence time is limited mainly by the hole population decay. We tentatively conclude that the main pure dephasing source of hole orbital states is acoustic phonons~\cite{Ramsay2010d}. The dephasing caused by the charge noise is negligible in our sample evidenced by the blinking free autocorrelation measurement and linewidth measurement with pure Lorentzian lineshape (see Supplementary Figs.~3 and~4). The precession frequency of $\left|\psi_1\right>$ can be extracted through the Fourier transform of the Ramsey fringes (Fig.~\ref{Ramsey}e). The peak at $\nu=$ 1.04 THz determined by the frequency difference of two orbital states well matches the value of 1.05(1) THz obtained from the fluorescence spectrum (Fig.~\ref{Observation}c). We note that the protocol demonstrated in Fig.~\ref{Ramsey} can be extended to prepare superposition states consisted of arbitrary hole orbital states (see Supplementary Fig.~9).

\begin{figure*}
\refstepcounter{fig}
	\includegraphics[width=0.85\linewidth]{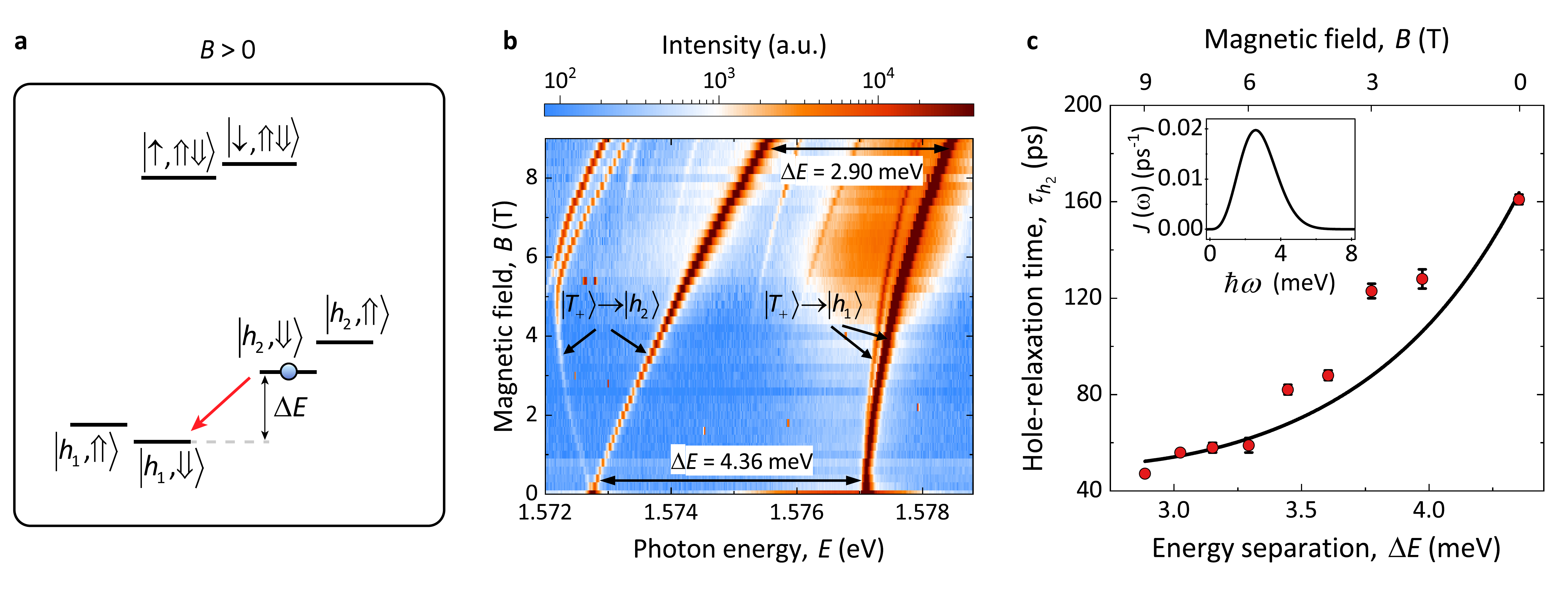}
	\caption{\textbf{Energy separation dependence of hole-relaxation time.} \textbf{a}, The energy-level diagram of the QD in a Faraday magnetic field. $\uparrow (\downarrow)$ and $\Uparrow (\Downarrow)$ represent electron and hole with spin up (down) respectively. Red arrow: phonon-assisted relaxation process. \textbf{b}, Fluorescence spectra under CW excitation with a magnetic field. Both the fundamental line ($\left|\textit{T}_\text{+}\right>\rightarrow \left|h_1\right>$) and the Auger peak ($\left|\textit{T}_\text{+}\right>\rightarrow \left|h_2\right>$) are divided into two peaks due to Zeeman splittings. The detuning $\Delta E$ between the higher energy branches of the fundamental line and the Auger peak is determined by the energy separation between $\left|h_1, \Downarrow\right>$ and $\left|h_2, \Downarrow\right>$. $\Delta E$ is tuned from 4.36 to 2.90~meV with the increase of the magnetic field due to different g-factors and diamagnetic shifts of $h_1$ and $h_2$ orbitals. \textbf{c}, The $\left|h_2,\Downarrow\right>$ to $\left|h_1,\Downarrow\right>$ relaxation time $\tau_{h_2}$ as a function of $\Delta E$. $\tau_{h_2}$ is measured according to the scheme shown in Fig.~\ref{decay}. Error bars arise from single-exponential fitting residual standard error. Black line: calculated hole-relaxation time. Inset: calculated spectral density of the exciton-LA phonon interaction $J(\omega)$.}
	\label{magnetic dep}
\end{figure*}

\addtocontents{toc}{\SkipTocEntry}
\section*{Single-hole relaxation dynamics}

The capability of coherently controlling a high-orbital charge carrier would enable a wide range of applications in, e.g. material science and quantum photonics. As an example, we apply this technique to the study of single-hole relaxation dynamics in a QD without the influence of additional carriers, which is previously rarely explored experimentally via optical excitation but crucial for optimizing QD-based light sources~\cite{Zibik2009,Tredicucci2009} and optical switches~\cite{Jin2009}. In QDs, single carriers relax from high orbitals mainly via emitting phonons~\cite{Vurgaftman1994a,Cogan2023}. The relaxation time is dependent on the density of phonon modes with energies matching hole-level separations. Since hole levels are densely spaced, hence easy to find longitudinal acoustic (LA) phonon modes with suitable energies, it is generally expected that, for InGaAs self-assembled QDs, the hole-relaxation time is negligible, namely tens of picoseconds or less~\cite{Vurgaftman1994a,Efros1995,Cogan2023}. However, for droplet-etched GaAs QDs, both short (29~ps~\cite{Hopfmann2021}) and long ($\sim$1.84 ~ns~\cite{Reindl2019b}) relaxation times of excitons with an excited hole are reported. The reason for the notably different relaxation times reported in the literature is still unclear. Furthermore, the presence of additional charge carriers may affect the hole relaxation dynamics via Coulomb interactions~\cite{Efros1995}, making it difficult to extract the relaxation time of a single excited hole. Here, with the method of deterministic generation of a single high-orbital hole, we find a relaxation time of 161(1)~ps with the second lowest hole state $h_2$ and around one orders of magnitude shorter relaxation time, e.g. 15(1)~ps, with even higher hole orbital states (see Fig.~\ref{decay}).

The hole-relaxation time is measured as follows. We use a pair of two-color $\pi$ pulses to deterministically prepare a high-orbital hole (see Figs.~\ref{decay}{a-d}) and then another $\pi$ pulse to probe the evolution of the high-orbital state population which is proportional to the $\left|\textit{T}_\text{+}\right>$ emission intensity (see Figs.~\ref{decay}{e-h}). During the interval $\Delta\textit{t}$ between the preparation and probe pulses, the population of the high-orbital state decays from 1 to $e^{-\Delta\textit{t}/\tau}$ via phonon-assisted relaxation. Figures~\ref{decay}i, j, k and l show the measured population of different orbital states as a function of $\Delta\textit{t}$. $\tau$ = 161(1), 28(1), 22(1), 15(1)~ps for $h_2$, $h_3$, $h_4$, $h_5$ are obtained respectively. We note that the hole-relaxation time of $h_2$ ($\tau_{h_2}$) is 6-10 times longer than $\tau_{h_3}$, $\tau_{h_4}$ and $\tau_{h_5}$. We attribute the relatively long $\tau_{h_2}$ to the phonon bottleneck effect in which the phonon-assisted relaxation is strongly inhibited because of the mismatch between hole level separations and phonon energies~\cite{Benisty1991,Urayama2001}. This interpretation is supported by the fact that the separation (4.36 meV) between $\left|h_2\right>$ and $\left|h_1\right>$ is clearly beyond the LA phonon sideband covering around 3~meV at the lower energy side of the fundamental line (see Fig.~\ref{Observation}c). By contrast, the separations (1-2~meV) between higher-orbital hole states are well within the LA phonon sideband. In order to understand whether a multi-phonon processes play a role here, we also measure the filling time (197(4)~ps) of $\left|h_1\right>$ under the condition where the QD is initially pumped to $\left|h_5\right>$ (see Supplementary Section~VI). The agreement between the experiment and a rate-equation simulation yielding a filling time of 173(1)~ps implies that the hole relaxation process studied here is dominated by a single-phonon-assisted cascade relaxation process. In addition to the energy difference, the hole relaxation time could also be affected by selection rules. A $\mu$s-long simulated hole tunneling time and the independence of $\tau$ on the bias safely excludes a contribution of hole tunneling from the measured $\tau$ (see Supplementary Section~VIII). The long-lived orbital state observed here is highly beneficial for orbital-encoded qubits and quantum cascade lasers~\cite{Zibik2009,Tredicucci2009}.

To further verify that the relatively long $\tau_{h_2}$ is indeed caused by the phonon bottleneck effect, we employ a magnetic field to continuously tune the $\left|h_2\right>$ to $\left|h_1\right>$ separation. At \textit{B} \textgreater 0, the spin degeneracy of $\left|h_1\right>$, $\left|h_2\right>$ and $\left|\textit{T}_\text{+}\right>$ is lifted (see Fig.~\ref{magnetic dep}a), resulting in a splitting of both the fundamental line ($\left|\textit{T}_\text{+}\right>\rightarrow \left|h_1\right>$) and the Auger peak ($\left|\textit{T}_\text{+}\right>\rightarrow \left|h_2\right>$) (see Fig.~\ref{magnetic dep}b). Here we focus on the hole relaxation dynamics between $\left|h_2, \Downarrow\right>$ and $\left|h_1, \Downarrow\right>$ whose energy separation $\Delta E$ is equal to the detuning between the higher-energy branches of the Auger peak and the fundamental line. With an increase of the magnetic field, $\Delta E$ decreases from 4.36 to 2.90~meV, leading to a more effective hole-phonon coupling. Consequently, $\tau_{h_2}$ measured according to the scheme shown in Fig.~\ref{decay} reduces from 161 to 47~ps (see Fig.~\ref{magnetic dep}b). This trend is consistent with the phonon-bottleneck interpretation and can be well-fitted by the theoretical model~\cite{Iles-Smith2017c,Madsen2013,Reiter2019} (see the solid line in Fig.~\ref{magnetic dep}c):

\begin{equation}
    J(\omega) = \sum_{\bf q} |\gamma_{\bf q}|^{2}\,\delta(\omega-\omega_{\bf q})=\alpha\omega^3\exp(-\omega^2/\omega_c^2).
\label{fermi_golden}
\end{equation}
Here $J(\omega)$ is the phonon spectral density of the hole-phonon interaction, which is proportional to phonon-assisted relaxation rate $1/\tau_{h_2}$. $\gamma_{\bf q}$ is the coupling strength to a phonon with wave-vector \textbf{q}, $\omega_c$ is the phonon cut-off frequency, and $\alpha$ is the scaling factor. Using $\omega_c$ and $\alpha$ as fitting parameters, we obtained good agreement with our experiment for $\hbar\omega_c=2.11(5)$ meV and $\alpha=0.0052(2)~\textrm{ps}^2$, which are similar to that in ref.~\onlinecite{Reigue2017a}. The corresponding spectral density is shown in the inset of Fig.~\ref{magnetic dep}c.

\addtocontents{toc}{\SkipTocEntry}
\section*{Discussion}
In addition to directly measuring single-carrier relaxation dynamics, the stimulated Auger process demonstrated in this work would enable various orbital-based applications. For example, quantum information can be encoded with hole orbital states. Compared with other types of qubits in QDs, orbital qubits may have several advantages. Compared to spin qubits~\cite{Press2008,DeGreve2011a,Greilich2011,Godden2012,Greve2013}, orbital qubits do not require strong magnetic fields. The fact that the coherence time is essentially limited by the hole lifetime suggests that it could be extended by fabricating devices with even larger energy splitting between $\left|h_1\right>$ and $\left|h_2\right>$. A hole lifetime as long as 100 ns~\cite{Zibik2009} could be achieved by controlling the QD growth conditions~\cite{Pan2000,Zibik2007}. Additionally, orbital and spin degrees of freedom could also be combined to realize a controlled-NOT gate in optically active solid-state QDs similar to that implemented with trapped ions~\cite{Monroe1995}. Compared to excitonic qubits composed of e.g. $\left|h_1\right>$ and $\left|T_+\right>$~\cite{Zrenner2002a,Li2003}, orbital qubits are in theory less affected by phonon-induced dephasing. This is because phonons react to changes in the charge distribution, which is much more prominent when adding an additional pair of charge carriers as opposed to having the single carrier in a state with a slightly different probability density.

\addtocontents{toc}{\SkipTocEntry}
\section*{Conclusions}
In summary, we demonstrate an optical scheme to coherently control a high-orbital hole via the stimulated Auger process with near-unity fidelity. The observation of Rabi oscillations and Ramsey interference proves the coherent nature of the Auger process. The coherence time is limited mainly by the hole relaxation time. This scheme simplifies the requirement to access orbital states with THz spacings, therefore opening a new avenue to gaining insight into the carrier dynamics of quantum emitters and to creating orbital-based quantum photonic devices. As an example, applying this scheme to a positively charged QD allows us to observe the phonon bottleneck effect and a single-phonon-assisted cascade relaxation process. 

Our results constitute a key step towards the study of a variety of quantum optical phenomena in solid-state multi-orbital level systems, such as stimulated Raman transitions between two orbital states and the generation of orbital-frequency entanglement between a charge carrier and a photon. Furthermore, pulse-driven stimulated Auger process can be used to depopulate the excited QDs with high speed and fidelity, providing a new mechanism to realize ultrafast all-optical switches~\cite{Sun2018,Jeannic2022} and super-resolution imaging~\cite{Kaldewey2018a,Kianinia2018}. In addition to QDs, our approach can be potentially extended to coherently controlling other quantum emitters, including two-dimensional materials~\cite{Chen2021c,Chow2020} and colloidal nanostructures~\cite{Antolinez2019}, where the Auger process plays a key role in the dynamics of exciton recombination.

\newpage
\addtocontents{toc}{\SkipTocEntry}
\section*{Acknowledgement}
We thank Nadine Viteritti for electrical contact preparation of the sample and Prof. Doris E. Reiter for fruitful discussions. F.L., D.-W.W. and W.C. acknowledge support by the National Natural Science Foundation of China (U21A6006, 62075194, 61975177, U20A20164, 11934011, 62122067) and Fundamental Research Funds for the Central Universities (2021QNA5006). H.-G.B., A.D.W., and A.L. acknowledge support by the BMBF-QR.X Project 16KISQ009 and the DFH/UFA, Project CDFA-05-06.
\addtocontents{toc}{\SkipTocEntry}
\section*{Author contributions}

F.L. and J.-Y.Y. conceived the project. H.-G.B., A.D.W. and A.L. grew the wafer and fabricated the sample. Y.-T.W., H.D and W.C. performed optical and electronic transport simulations for the sample. J.-Y.Y., C.C., X.-D.Z. and Y.M. carried out the experiments. J.-Y.Y., M.C. and F.L. analysed the data. J.-Y.Y and D.-W.W. performed the quantum dynamics simulation. X.H., W.F., X.L., D.-W.W., C.-Y.J. and F.L. provided supervision and expertise. J.-Y.Y. and F.L. wrote the manuscript with comments and inputs from all the authors.

\addtocontents{toc}{\SkipTocEntry}
\section*{Competing Interests}
The authors declare no competing interests.

\addtocontents{toc}{\SkipTocEntry}
\clearpage

\addtocontents{toc}{\SkipTocEntry}

\end{bibunit}

\newpage
\begin{bibunit}[naturemag]

\addtocontents{toc}{\SkipTocEntry}
\section*{Methods} 

\label{sec:methods}

\addtocontents{toc}{\SkipTocEntry}
\subsection*{Sample preparation}

Sample growth was performed following the general heterostructure design previously published~\cite{Babin2021,Zhai2020}. We adapted the growth temperatures and modified the local droplet etching (LDE) process. The heterostructure layers were grown at a pyrometer reading of 605 °C unless stated otherwise. Modifications to the LDE process relevant to QD formation were made. In a 240 s growth interrupt, an LDE etch process temperature of 560 °C and an arsenic beam equivalent pressure (BEP) of $4.5\times10^{-7}$ Torr are reached. This is followed by nominally 0.31 nm Al deposition and 120 s etch time. In this way, we obtain a monomodal and low ensemble inhomogeneity QD energy distribution. Moreover, Al and GaAs materials were deposited as gradients by sample rotation stop similar to sample A as detailed in ref.~\cite{Babin2022}. This gradient procedure yields approximately 70\% of the wafer with QDs. Thereafter As-flux was restored to match a BEP of $9.6\times10^{-6}$ Torr and the QDs were filled with 1.0 nm of GaAs, also in a gradient similar to ref.~\cite{Babin2022}. After an annealing break of 180 s, the QDs were capped with $\rm Al_{0.33}Ga_{0.67}As$ and sample growth was continued as described in ref.~\cite{Babin2021}. 
For sample processing, photolithographic etching and contact masks were used. Mesas by deep etching and n-contact windows by removing the p-layer with sulfuric acid, hydrogen peroxide, and water etch (1:1:50) were made. Then, n-contacts consisting of 10 nm Ni, 60 nm Ge, 120 nm Au, 10 nm Ni and 100 nm Au were deposited in high vacuum by thermal evaporation and annealed under forming gas at 420 °C for 120 s. Thereafter the p-contacts were deposited by depositing 10 nm Au, 15 nm Cr, and 200 nm Au without further annealing. 

\addtocontents{toc}{\SkipTocEntry}
\subsection*{Measurement techniques}
The schematic of experimental setup is presented in Extended Data Fig.~1. The QD device was cooled to 3.6 K in a cryostat equipped with a superconducting magnet. Sample excitation and collection were performed with a home-built confocal microscope. The applied electrical bias was set near the center of the single-hole-charging voltage plateau (see Supplementary Fig.~1). The collected fluorescence signal was sent to a 750-mm monochromator and then detected on a Si charge-coupled device detector. The suppression of the resonant excitation laser was ensured by a pair of polarizers working at a cross-polarization configuration~\cite{Kuhlmann2013b}. For CW excitation, a tunable narrow-linewidth (full width at half maximum \textless 100 kHz measured over a period of 100 $\mu s$) Ti:sapphire laser was used. For pulsed excitation, a tunable mode-locked Ti:sapphire laser generated pulses with a pulse width of 140 fs at a repetition rate of 80 MHz. Two folded pulse shapers~\cite{Yan2022S} were used to pick out 6-ps duration phase-locked pulses with different colors operated as pump and control pulse (see detail in Extended Data Fig.~1). The time delay between these pulses was introduced by a motorized delay line. For Ramsey interference, the control laser path was further divided into two arms with another delay line. The temporal resolution of pump-probe technique is $\sim$9.5 ps, which is ultimately limited by the laser pulse duration (Supplementary Section~VI).

\addtocontents{toc}{\SkipTocEntry}
\section*{Data availability}

The raw data that support the findings of this study are available at \textcolor{blue}{https://doi.org/10.5281/zenodo.7947362} and from the corresponding author upon reasonable request.
\\

\addtocontents{toc}{\SkipTocEntry}
\section*{Code availability}
The codes that have been used for this study are available from the corresponding author upon reasonable request.

\addtocontents{toc}{\SkipTocEntry}
\section*{Additional information}
Supplementary information is available in the online version of this paper. Correspondence and requests for materials should be addressed to F.L.

\addtocontents{toc}{\SkipTocEntry}

\end{bibunit}

\setcounter{figure}{0}
\renewcommand{\figurename}{Extended Figure}
\clearpage
\newpage
\begin{figure*}
\refstepcounter{fig}
	\includegraphics[width=0.9\linewidth]{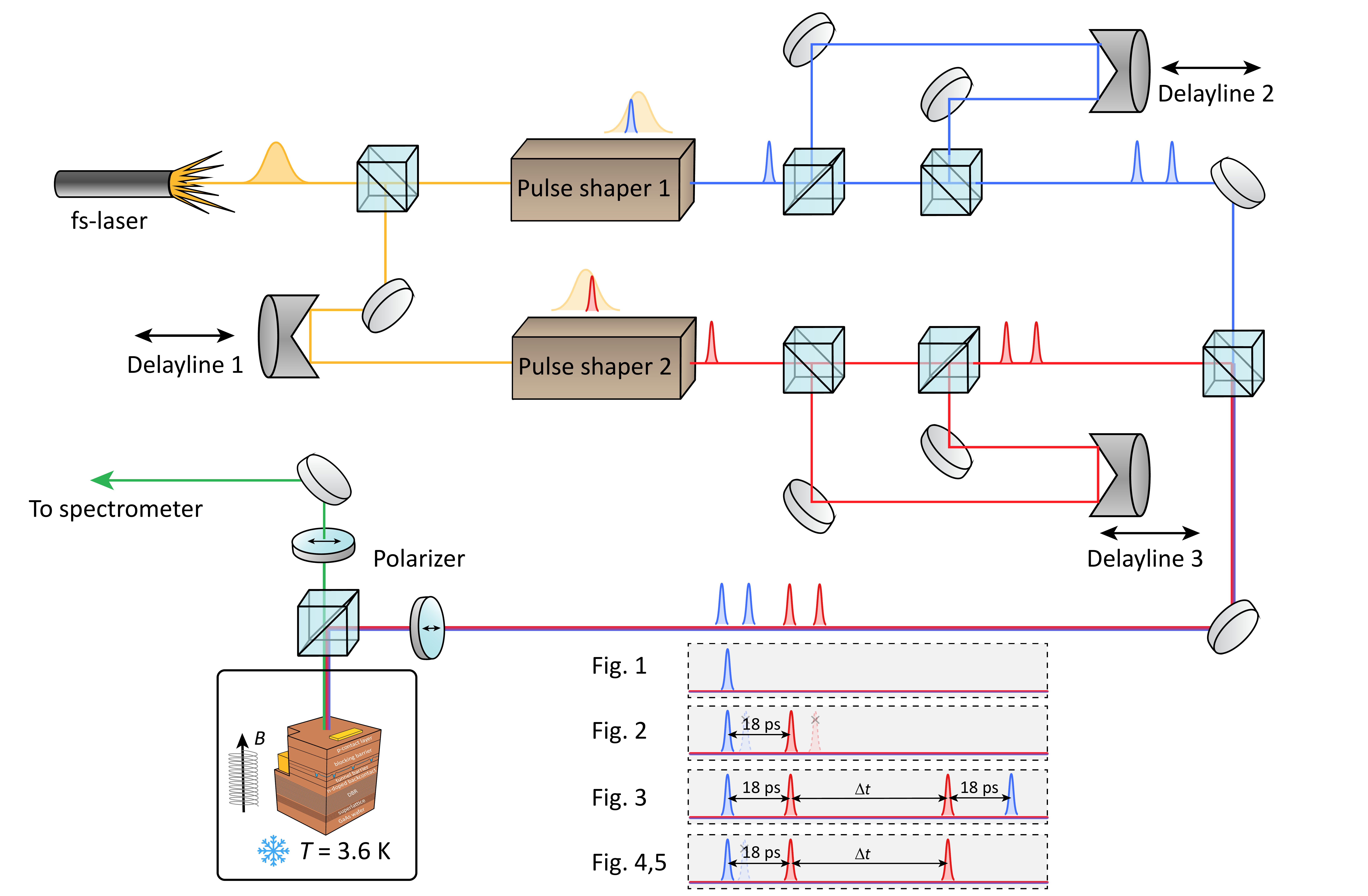}
	\caption{\textbf{Schematic of the experimental setup.} A single QD device is kept in a cryostat at 3.6 K and is excited under pulsed resonant excitation. A Ti-sapphire laser is used to generate 140 femtosecond optical pulses with an 80 MHz repetition rate. The femtosecond pulses are then sent into two folded pulse shapers to pick out two picosecond pulses with time delay controlled by delayline 1. Each pulse is divided into two arms and another controlled delay is introduced by delayline 2 and 3. Finally, the four pulses are recombined and focused on the sample by an NA=0.81 objective lens. A pair of polarizers working at a cross-polarization configuration is used to filter out the resonant excitation laser. The experimental data shown in the main text are acquired with the pulse sequences shown in the lower right corner.}
	\label{setup}
\end{figure*}

\begin{bibunit}[naturemag]

\setcounter{equation}{0}
\setcounter{figure}{0}
\setcounter{table}{0}

\renewcommand\thesection{\Roman{section}}
\renewcommand{\figurename}{Supplementary Figure}
\renewcommand{\tablename}{Supplementary Table}
\renewcommand{\thetable}{\arabic{table}}

\renewcommand{\theequation}{S\arabic{equation}}
\renewcommand{\bibnumfmt}[1]{[S#1]} 
\renewcommand{\citenumfont}[1]{S#1} 

\onecolumngrid
\clearpage
\newpage

\begin{Large}
\begin{center}
\textbf{Supplementary Information: Coherent control of a high-orbital hole in a semiconductor quantum dot}
\end{center}
\end{Large}

\setcounter{page}{1}

\begin{center}

Jun-Yong Yan, Chen Chen, Xiao-Dong Zhang, Yu-Tong Wang, Hans-Georg Babin,\\
Andreas D. Wieck, Arne Ludwig, Yun Meng, Xiaolong Hu, Huali Duan, Wenchao Chen,\\
Wei Fang, Moritz Cygorek, Xing Lin, Da-Wei Wang, Chao-Yuan Jin, and Feng Liu
\end{center}

\tableofcontents

\newpage
\section{Identification of the positively charged state}
\begin{figure}[h]
\includegraphics[width=1\textwidth]{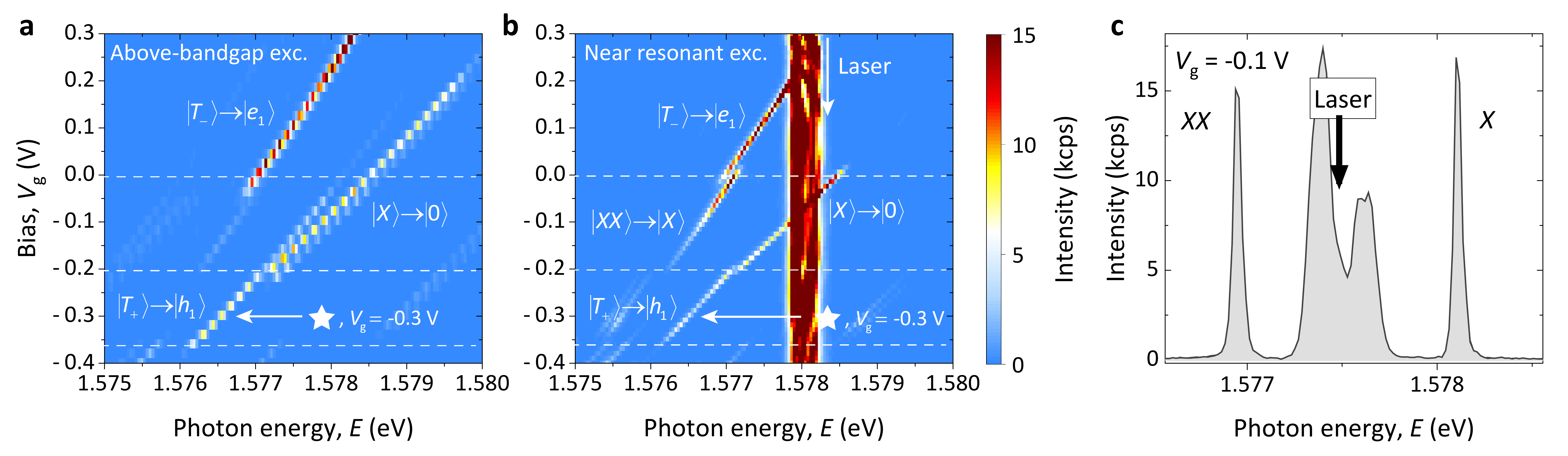}
\caption{Tuning the charge state of the QD. \textbf{a}, Bias-dependent fluorescence spectra with above-bandgap excitation. Photon energy of the excitation laser: 1.946 eV. The horizontal dashed lines mark the charging events. The white arrow indicates the bias of $\textit{V}_g$ = -0.3~V where we performed all the measurements in the main text. \textbf{b}, Bias-dependent fluorescence spectra with near-resonant pulsed excitation. Photon energy of the excitation laser: 1.578 eV (indicated by white vertical arrow). \textbf{c}, The spectrum of the emitted \textit{XX} and \textit{X} photons under TPE at $\textit{V}_g = -0.1$~V.}
\label{fig:Bias map}
\end{figure}

To make sure that our QD is positively charged, we measure the bias-dependent fluorescence spectra under above-bandgap excitation (Supplementary Fig.~\ref{fig:Bias map}a) and near-resonant excitation (Supplementary Fig.~\ref{fig:Bias map}b), showing distinct Coulomb
blockade with a series of charge plateaus. The neutral exciton (\textit{X}) and biexciton (\textit{XX}) states are identified by observing the two-photon resonant excitation (TPE) of the biexciton~\cite{Stufler2006a} as shown in Supplementary Fig.~\ref{fig:Bias map}c. The assignment of other charge states can be deduced from the band structure calculation at different bias (see Supplementary Fig.~\ref{fig:wkb}a).

The QD is positively charged via electron tunneling (see Supplementary Fig.~\ref{fig:electron tunnel}). At $\textit{V}_g=-0.3$ V, the QD is initially at crystal ground state ($\left|0\right>$). When the pulse resonant with $\left|h_1\right>\leftrightarrow\left|\textit{T}_\text{+}\right>$ transition is turned on, transition $\left|0\right>\leftrightarrow\left|\textit{X}\right>$ can also be driven~\cite{Tomm2021b,Delteil2016}. Then the electron in $\left|\textit{X}\right>$ tunnels out rapidly, leaving behind a long-lived (at least 100 $\mu$s, see Supplementary Fig.~\ref{fig:g2}) hole in the QD. Finally, the resonant pulse drives the $\left|h_1\right>\leftrightarrow\left|\textit{T}_\text{+}\right>$ transition,  preparing a positive trion $\left|\textit{T}_\text{+}\right>$.

\begin{figure}[h]
\includegraphics[width=0.25\textwidth]{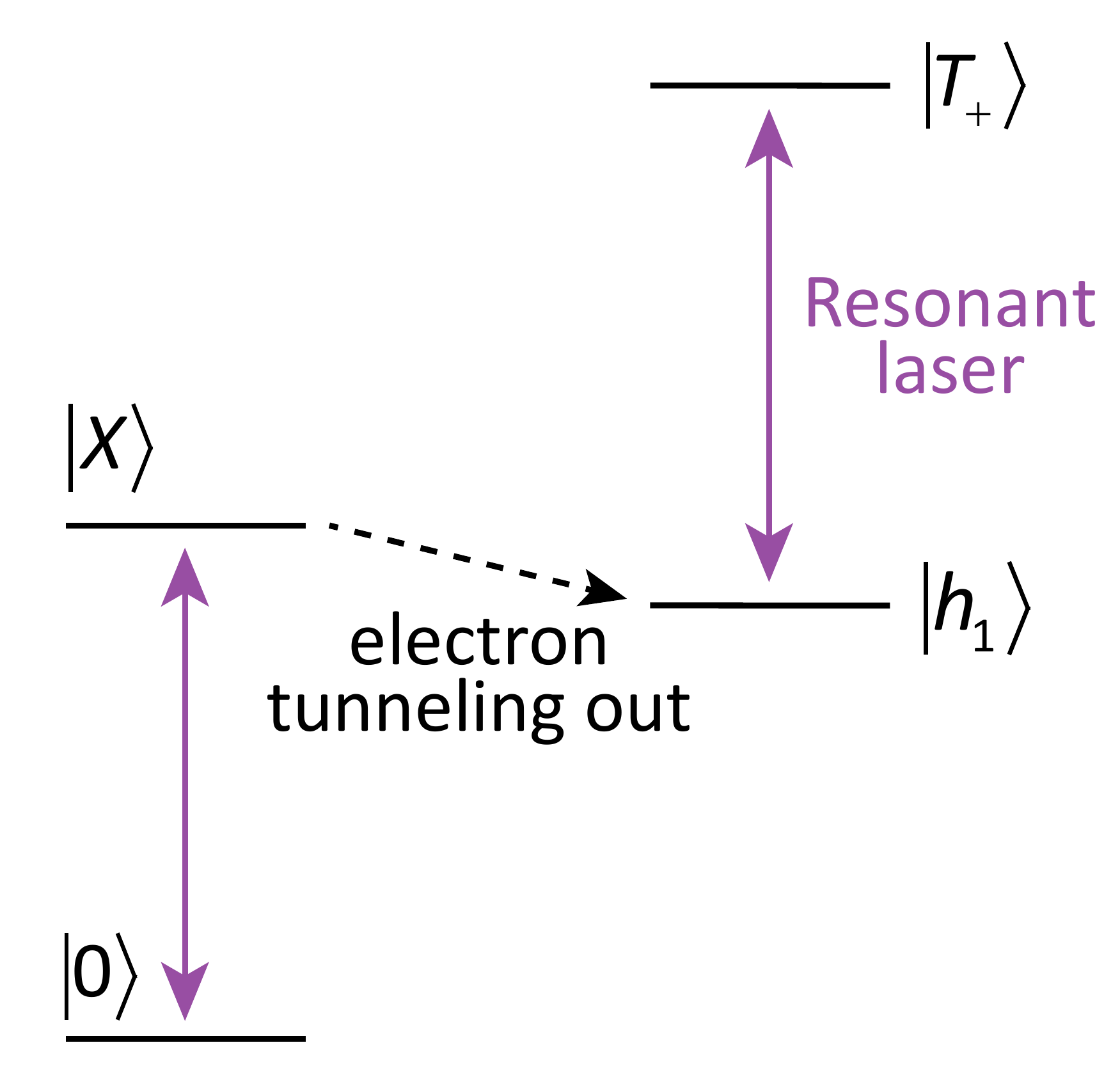}
\caption{Optical charging via electron tunneling.}
\label{fig:electron tunnel}
\end{figure}

To evaluate the stability of $\left|\textit{T}_\text{+}\right>$, we measure the intensity-correlation histogram of the $\left|\textit{T}_\text{+}\right>\rightarrow\left|h_1\right>$ emission using a Hanbury Brown and Twiss-type setup. The long-timescale intensity-correlation histogram is shown in Supplementary Fig.~\ref{fig:g2}. The heights of side-peaks in $g^2$ measurement all stay quite close to unity (see the orange dotted line), indicating the absence of blinking~\cite{Zhai2022} and, hence, a long-lived hole in $h_1$ orbital. The finite $g^2(0)$ comes likely from the residual laser scattering and re-excitation process~\cite{Fischer2017a,Loredo2019}. 

\newpage
\begin{figure}[h]
\includegraphics[width=1\textwidth]{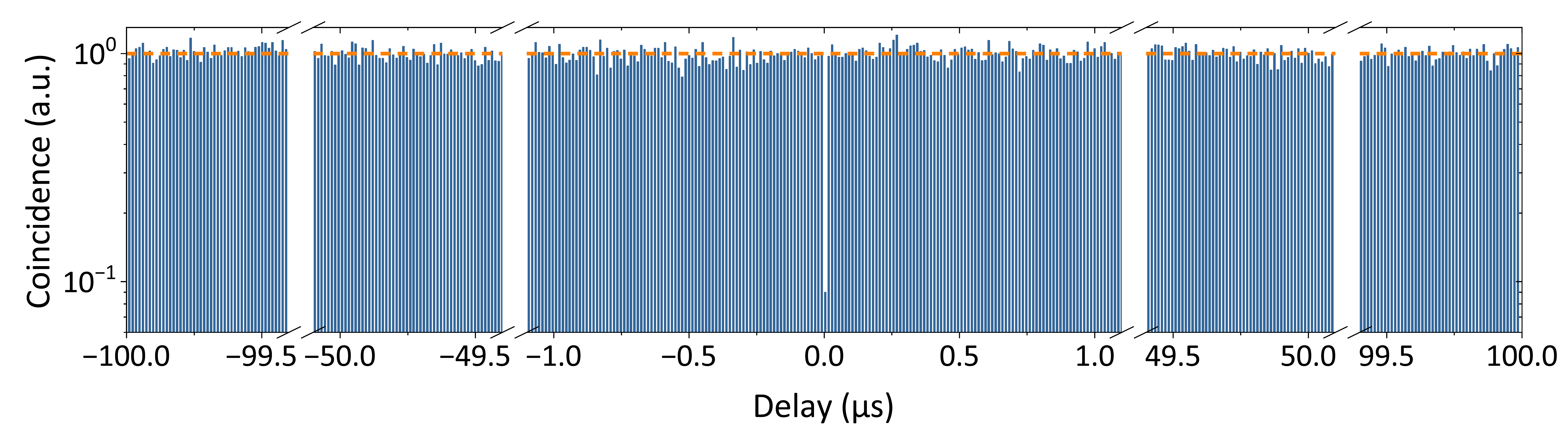}
\caption{Intensity-correlation histogram of the $\left|\textit{T}_\text{+}\right>\rightarrow\left|h_1\right>$
emission from the QD under pulsed excitation resonant with the $\left|h_1\right>\leftrightarrow\left|\textit{T}_\text{+}\right>$ transition measured using a Hanbury Brown and Twiss-type setup.}
\label{fig:g2}
\end{figure}

\begin{figure}[h]
\includegraphics[width=0.3\textwidth]{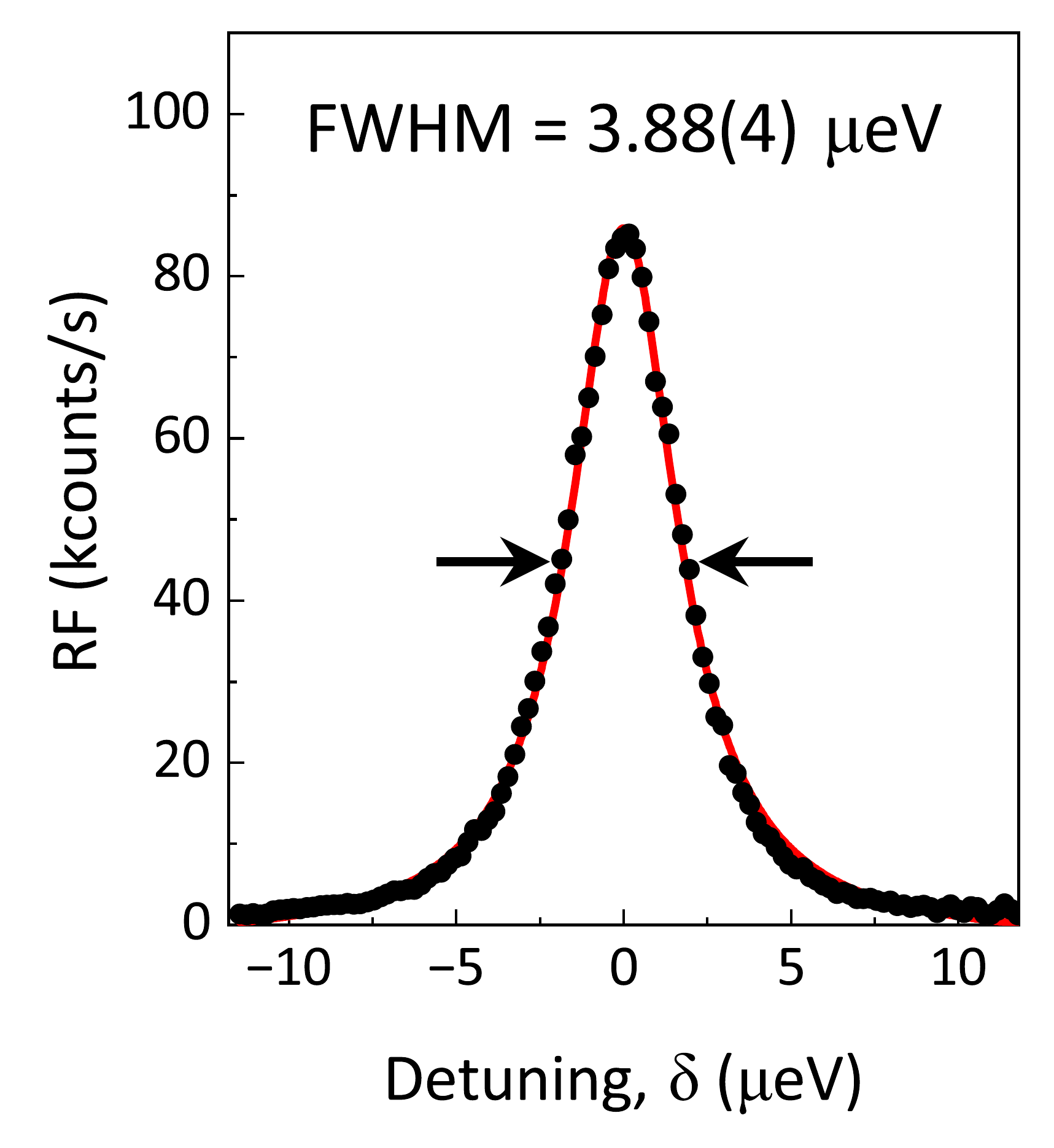}
\caption{The linewidth of $\left|\textit{T}_\text{+}\right>$ fundamental transition measured by scanning a narrow-bandwidth CW laser across the $T_+$ resonance. Red curve: a fitting using a Lorentzian function.}
\label{fig:cw scan linewidth}
\end{figure}

In addition, a linewidth of 3.88(4) $\mu$eV was obtained by scanning a narrow-bandwidth CW laser across the $T_+$ fundamental transition (Supplementary Fig.~\ref{fig:cw scan linewidth}). The resonance fluorescence (RF) data can be well fitted by a Lorentzian function (red curve), indicating an absence of inhomogeneous broadening caused by a fluctuating charge environment~\cite{Kuhlmann2013d}.

\newpage
\section{Magnetic-field-dependent fluorescence spectra}
\label{sec:sample}

\begin{figure}[h]
\refstepcounter{fig}
	\includegraphics[width=0.7\linewidth]{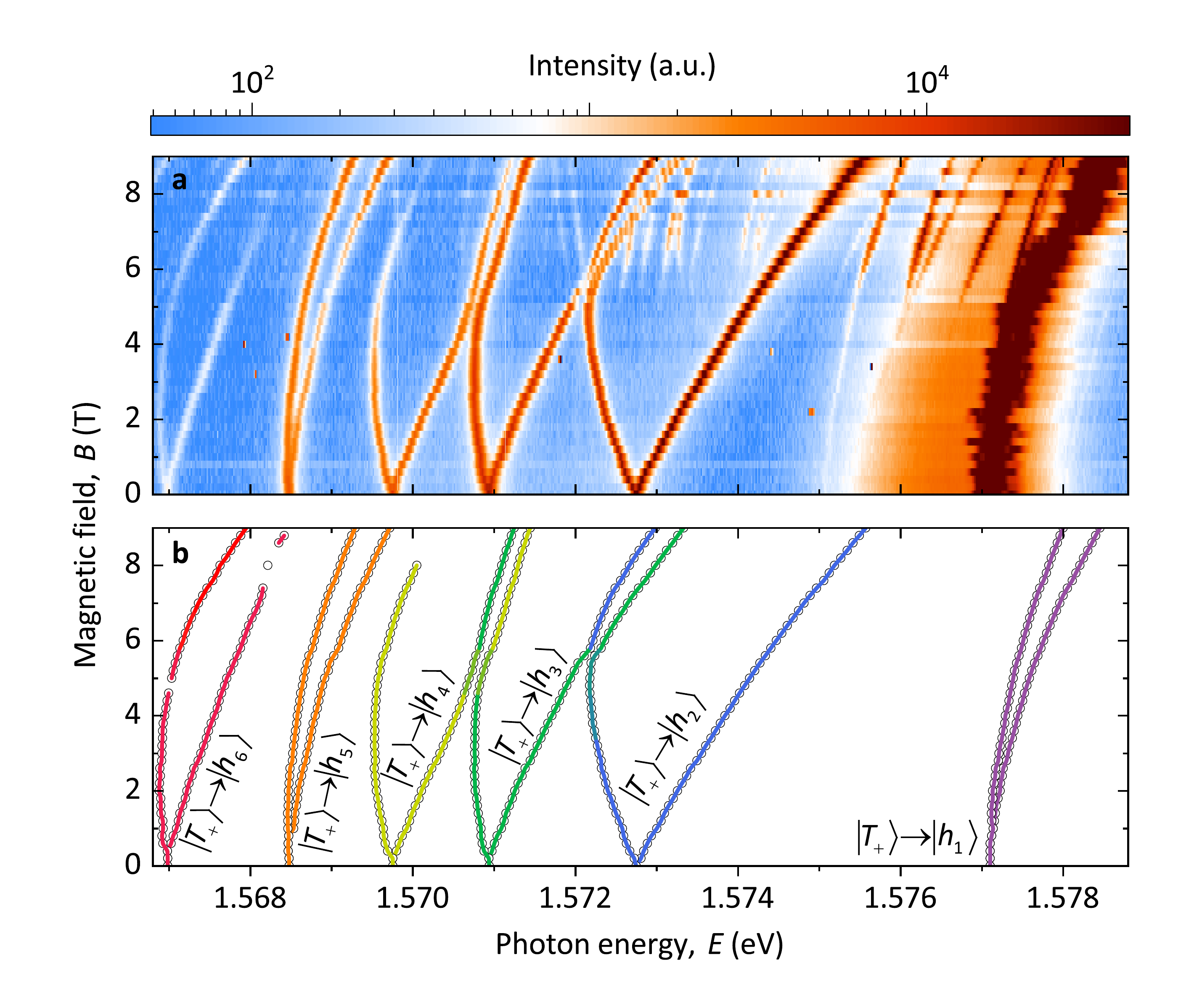} 
	\caption{\textbf{a}, Magneto-optical spectra under pulsed excitation resonant with the $\left|h_1\right>\leftrightarrow\left|\textit{T}_\text{+}\right>$ transition. Two anti-crossing points are observed indicating the energy degeneration points of single-hole orbital levels. \textbf{b}, Central energy of fluorescence peaks extracted by fitting each peak in \textbf{a} with a Gaussian function.}

\label{sfig:Magnetic}
\end{figure}

To gain more insight into radiative Auger transitions, we apply a Faraday magnetic field to the sample. Supplementary Fig.~\ref{sfig:Magnetic} shows the fluorescence spectra as a function of magnetic ﬁeld. The external field gives rise to a pair of Zeeman-split fluorescence peaks for the fundamental and Auger transitions. Two anti-crossing points of Auger peaks are observed around \textit{B} = 5~T, indicating the resonance of hole orbital states~\cite{Blokland2007,Climente2005,Reuter2005}. This feature is a strong evidence to distinguish the Auger peaks from emission of other charged states, phonon-assisted transitions or nearby quantum dots.

\begin{figure}[h]
\refstepcounter{fig}
	\includegraphics[width=0.7\linewidth]{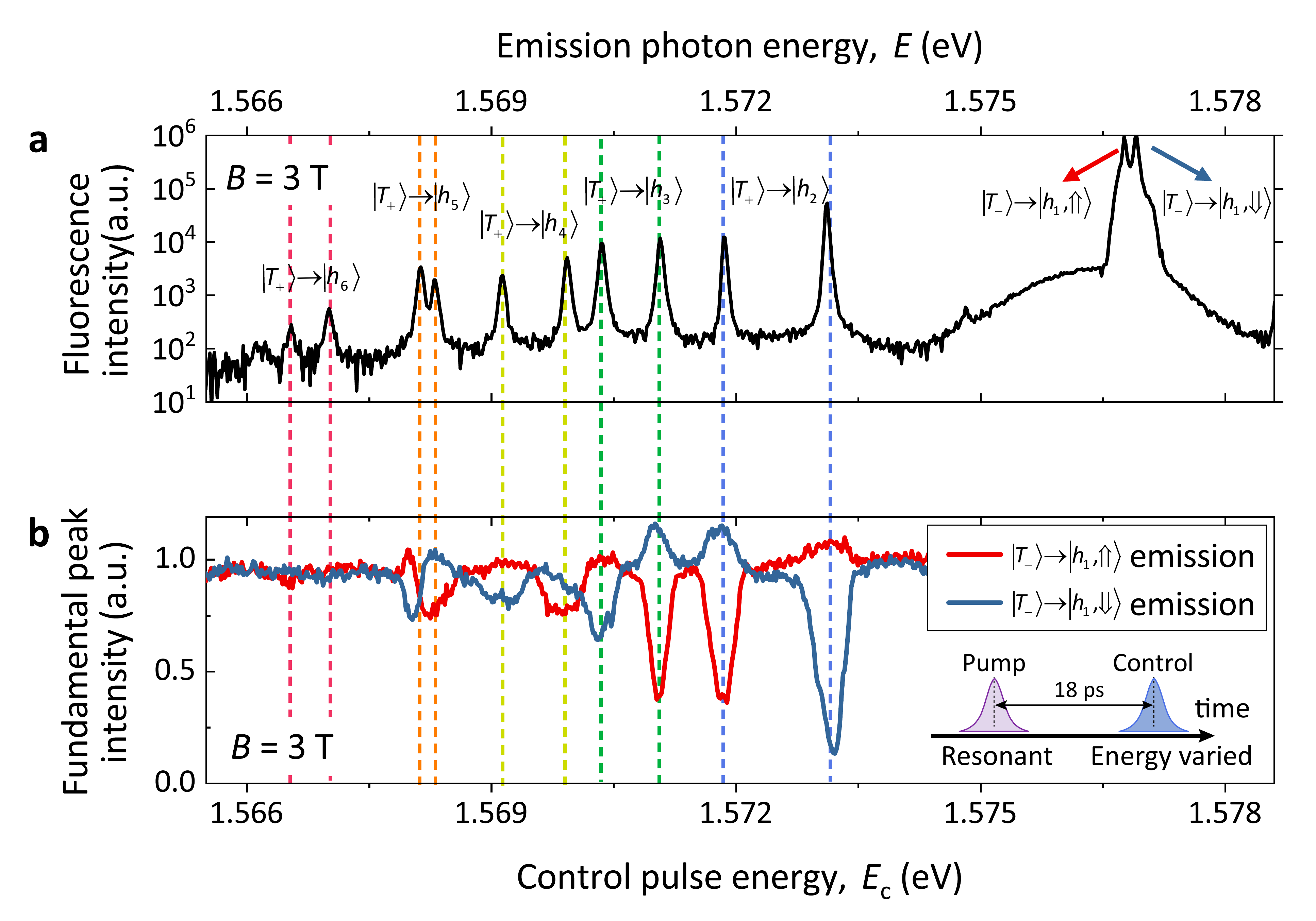}
	\caption{\textbf{a}, Fluorescence spectrum obtained by pulsed excitation resonant with the $\left|h_1\right>\leftrightarrow\left|\textit{T}_\text{+}\right>$ transition at \textit{\textbf{B}} = 3~T. The dashed lines serve as a guide to the eye. \textbf{b}, Dependence of two fundamental lines on the control-pulse energy obtained by tuning the control-pulse energy while recording the intensity of two fundamental emission lines (red curve: $\left|T_\textrm{+}\right>\rightarrow\left|h_1,\Uparrow\right>$, blue curve: $\left|T_\textrm{+}\right>\rightarrow\left|h_1,\Downarrow\right>$). Inset: a timing diagram of the pulse sequence.}

\label{sfig:depopulation at 3T}
\end{figure}

To clarify the selection rule of coherent Auger transitions, we measure the dependence of two fundamental lines on de-excitation (control) pulse energy at \textit{B} = 3~T.  Supplementary Fig.~\ref{sfig:depopulation at 3T}a presents a fluorescence spectrum at \textit{B} = 3~T without the control pulse. The fundamental line and Auger peaks are divided into two peaks due to the Zeeman splittings. Supplementary Fig.~\ref{sfig:depopulation at 3T}b shows the result of the de-excitation experiment which is conducted as follows. Firstly, a resonant pump pulse creates a $\left|\textit{T}_\text{+}\right>$, from which the resonance fluorescence or Auger photons are emitted as 
shown in Supplementary Fig.~\ref{sfig:depopulation at 3T}a. Secondly, the control pulse is applied and we monitor the emission intensity from the two fundamental peaks while varying the energy of the control pulse. The intensity of each fundamental peak, shown by the red and blue curves respectively, can only be quenched when the control pulse is in resonance with one of the two split Auger peaks. These results indicate a clear selection rule of stimulated Auger transitions and their spin-preserving nature.

\newpage

\newpage
\section{Time-resolved resonance fluorescence}
To measure the lifetime of $\left|\textit{T}_\text{+}\right>$, a time-correlated single photon counting card synchronized with the laser pulse train records the arrival times of individual photons detected by a superconducting nanowire single photon detector (SNSPD). Pumping the $\left|h_1\right>\leftrightarrow\left|\textit{T}_\text{+}\right>$ transition with a $\pi$ pulse, we measure a $\left|\textit{T}_\text{+}\right>$ lifetime of 400(2) ps (see Supplementary Fig.~\ref{fig:TRPL}). 

\begin{figure}[h]
\includegraphics[width=0.4\textwidth]{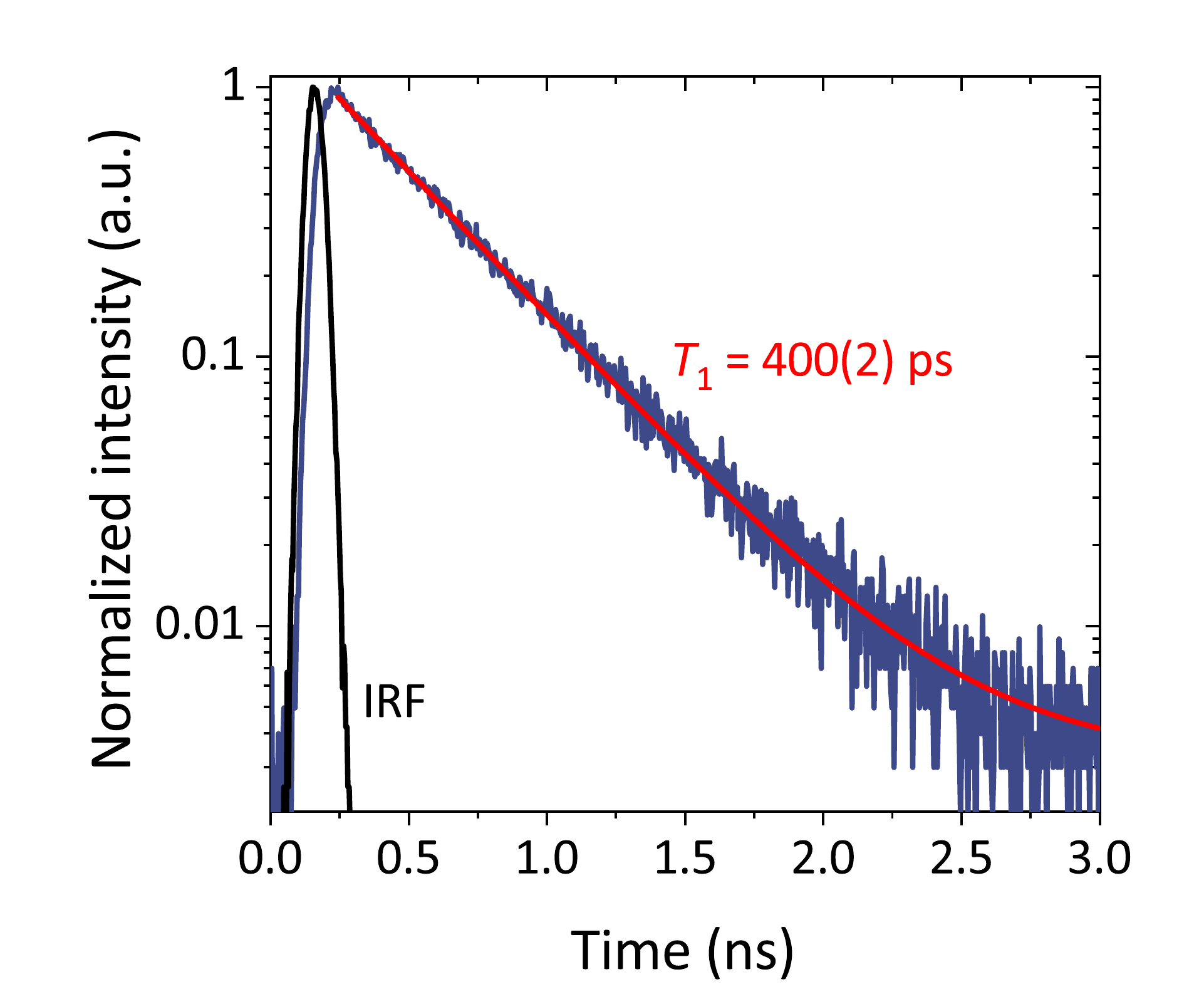}
\caption{Spontaneous emission lifetime of $\left|\textit{T}_\text{+}\right>$ measured with an SNSPD (full width at half maximum (FWHM) of the instrument response function $= 69.1(2)$~ps). Black peak: instruments response function (IRF). Violet curve: resonance fluorescence decay curve. Fitting the resonance fluorescence decay curve with a single exponential function (red line) yields the $\left|\textit{T}_\text{+}\right>$ lifetime $\textit{T}_\text{1}$ = 400(2)~ps.}

\label{fig:TRPL}
\end{figure}

\newpage
\section{Master-equation simulation}
\label{sec:Master}

To simulate the Rabi-oscillation results in Fig.~\ref{Rabi}, we consider a model for the three-level $\Lambda$ system coherently driven by two laser pulses. Due to the same polarization selection rules of the two transitions at $B=0$, each optical
ﬁeld couples to both dipole transitions~\cite{Steck2022} (cross-coupling, see Supplementary Fig.~\ref{Master simulation}c). The Hamiltonian with basis $\left|h_1\right>$, $\left|\textit{T}_{+}\right>$ and $\left|h_2\right>$ in the rotating frame for the above model takes the form:
\begin{equation}
\textit{H}_{eff.} = \hbar
\left( {\begin{array}{ccc}
0 & \frac{1}{2}\Omega_1+\frac{1}{2}\Omega'_2e^{-i(\Delta_{12}+\delta)t} & 0\\
\frac{1}{2}\Omega_1+\frac{1}{2}\Omega'_2e^{i(\Delta_{12}+\delta)t} & \Delta & \frac{1}{2}\Omega_2+\frac{1}{2}\Omega'_1e^{-i(\Delta_{12}+\delta)t} \\
0 & \frac{1}{2}\Omega_2+\frac{1}{2}\Omega'_1e^{i(\Delta_{12}+\delta)t} & -\delta
\end{array}}\right).
\label{Eq: ME}
\end{equation}

$\Delta$ is the detuning of the pump laser from $\left|\textit{T}_{+}\right>\leftrightarrow\left|h_1\right>$ transition, which is zero in our experiment. $\Delta+\delta$ is the detuning of the control laser from $\left|\textit{T}_{+}\right>\leftrightarrow\left|h_2\right>$ transition, which is swept to find the optimal control fidelity. $\hbar\Delta_{12}=4.36$ ~meV is the orbital energy splitting between $h_1$ and $h_2$. The QD-light interaction is speciﬁed by the Rabi frequency $\Omega_1=\mu_1\textit{E}_1/\hbar$, $\Omega_2=\mu_2\textit{E}_2/\hbar$, 
$\Omega'_{1}=\mu_2\textit{E}_1/\hbar$, and $\Omega'_{2}=\mu_1\textit{E}_2/\hbar$, where $\mu_1~(\mu_2)$ is the dipole moment of $\left|\textit{T}_{+}\right>\leftrightarrow\left|h_1\right>$ ($\left|\textit{T}_{+}\right>\leftrightarrow\left|h_2\right>$) transition and $\textit{E}_1~(\textit{E}_2)$ is the field amplitude of pump (control) pulse, respectively. Because no Purcell effect is involved here, we deduce the dipole moment ratio $\mu_1/\mu_2=5$ from the fluorescence intensity (see Fig.~\ref{Observation} in the main text). To evaluate the $\left|\textit{T}_{+}\right>$ population after the control pulse, that corresponds to the element $\rho_{22}(\textit{t})$ of the $3\times3$ density matrix $\rho(\textit{t})$, we solve numerically the master equation with the help of the Quantum Toolbox in Python (QuTiP)~\cite{Johansson2012a}:
\begin{equation}
i\hbar\frac{d\rho(\textit{t})}{d\textit{t}}=[\textit{H}_{eff.}, \rho(\textit{t})].
\label{Eq: ME1}
\end{equation}

Supplementary Fig.~\ref{Master simulation}b (d) shows the simulated final $\left|\textit{T}_{+}\right>$ population as a function of the control pulse detuning $\delta$ and pulse area $\Theta$ without (with) cross-coupling. Supplementary Fig.~\ref{Master simulation}e is the experimental observation for the same pulse detuning range as in Supplementary Fig.~\ref{Master simulation}d. The cross-coupling shifts the QD resonant energy via alternating current (AC) Stark effect~\cite{Unold2004,Jundt2008}. To achieve efficient population transfer, the Stark shift must be compensated. Therefore, both theoretically and experimentally, the highest transfer efficiency has been observed at slightly detuned energies, which is more pronounced for larger dipole moment ratios, i.e., Auger transitions to higher orbital states. The solid lines (dotted lines) shown in Supplementary Figs.~\ref{Master simulation}f-i are the simulated Rabi oscillations with (without) considering cross-coupling, when the control pulse couples $\left|\textit{T}_{+}\right>\rightarrow h_2$ (\textbf{f}), $\left|\textit{T}_{+}\right>\rightarrow h_3$ (\textbf{g}), $\left|\textit{T}_{+}\right>\rightarrow h_4$ (\textbf{h}), and $\left|\textit{T}_{+}\right>\rightarrow h_5$ (\textbf{i}), respectively. The aperiodic Rabi oscillations are in agreement with the experimental data shown in Fig.~\ref{Rabi}g-i. 

\newpage
\begin{figure}[h]
\includegraphics[width=1\textwidth]{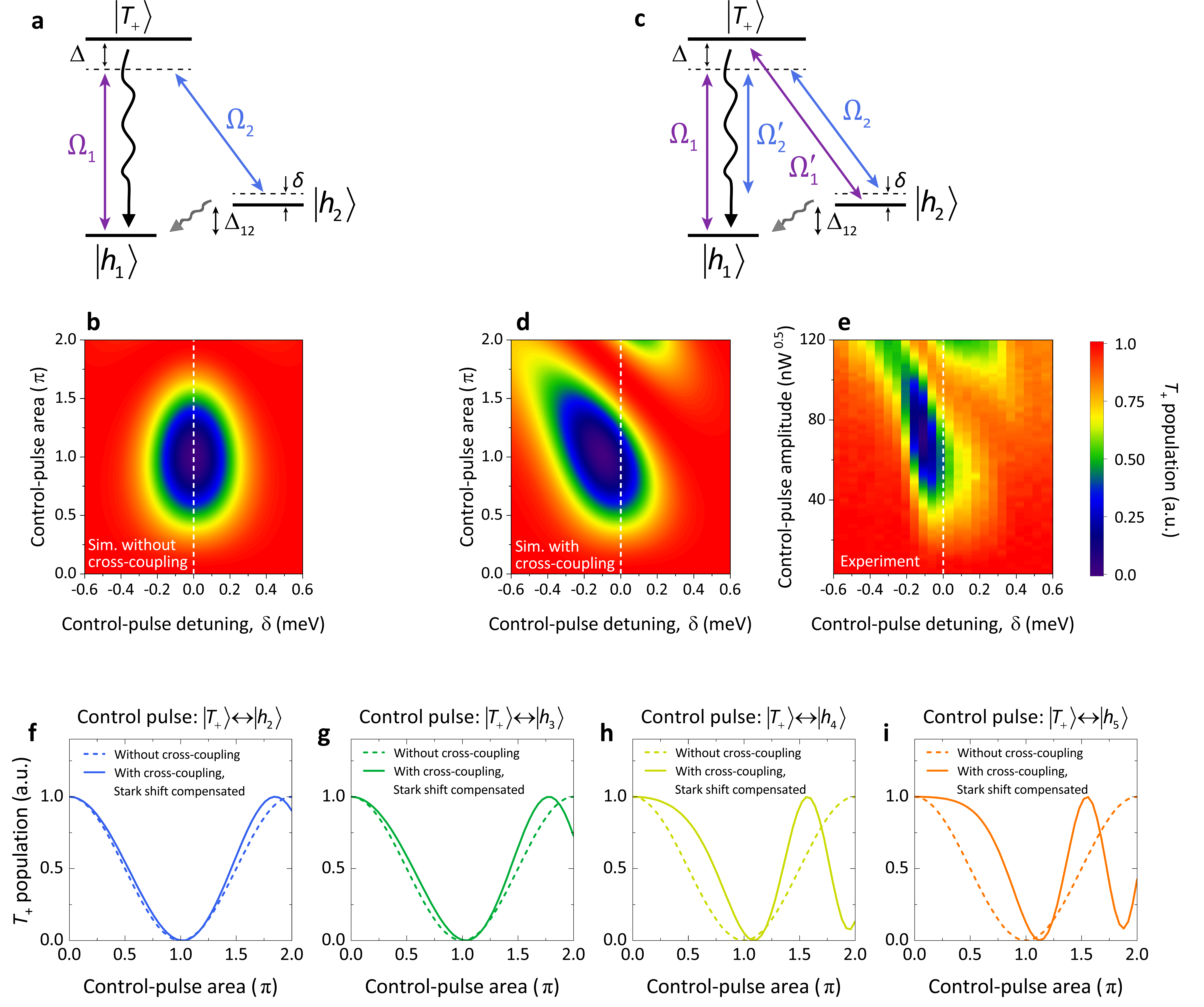}
\caption{Master-equation simulations of a $\Lambda$ system. \textbf{a (c)}, Schematic picture of the theoretical model without (with) considering cross-coupling. The violet (blue) arrow indicates the pump (control) laser ﬁeld. The black (gray) wavy arrow indicates the population decay of $\left|\textit{T}_{+}\right>$ ($\left|h_2\right>$) via spontaneous emission (phonon-assisted relaxation). \textbf{b (d)}, The final $\left|\textit{T}_{+}\right>$ population versus control pulse detuning $\delta$ and pulse area $\Theta$ without (with) cross-coupling. \textbf{e}, Experimentally obtained $\left|\textit{T}_{+}\right>$  population versus control-pulse detuning $\delta$ and pulse area $\Theta$. \textbf{e-i}, Simulated $T_+$ population as a function of control-pulse area with and without considering cross-coupling, when the control pulse couples $\left|\textit{T}_{+}\right>\rightarrow h_2$ (\textbf{f}), $\left|\textit{T}_{+}\right>\rightarrow h_3$ (\textbf{g}), $\left|\textit{T}_{+}\right>\rightarrow h_4$ (\textbf{h}), and $\left|\textit{T}_{+}\right>\rightarrow h_5$ (\textbf{i}), respectively. The rabi oscillations are aperiodic after considering the cross-coupling and compensating the Stark shift.}
\label{Master simulation}
\end{figure}

\newpage
\section{Protocol to prepare the superposition of arbitary hole orbital states}
\label{sec:superposition state}
Our protocol demonstrated in Fig.~\ref{Ramsey} is not limited to generating superposition states between the two lowest hole orbital states, but can be extended to prepare superposition states consisting of arbitrary hole orbital states. As shown in Supplementary Fig.~\ref{Fig:superposition state}, the creation of a superposition state involving multiple hole orbital states can be realized by first pumping the QD from $\left|h_1\right>$ to $\left|T_+\right>$  with a resonant pulse (indicated by the purple arrow) via the fundamental transition, followed by transferring the population of $\left|T_+\right>$  to different hole orbital states with a series of resonant pulses (indicated by blue, green... and red arrows) via stimulated Auger processes. The resulting superposition state can be expressed as $\left|\Psi\right>=a_1e^{i\phi_1}\left|h_1\right>+a_2e^{i\phi_2}\left|h_2\right>+a_3e^{i\phi_3}\left|h_3\right>+...+a_ne^{i\phi_n}\left|h_n\right>$.

\begin{figure}[h]
\includegraphics[width=0.35\textwidth]{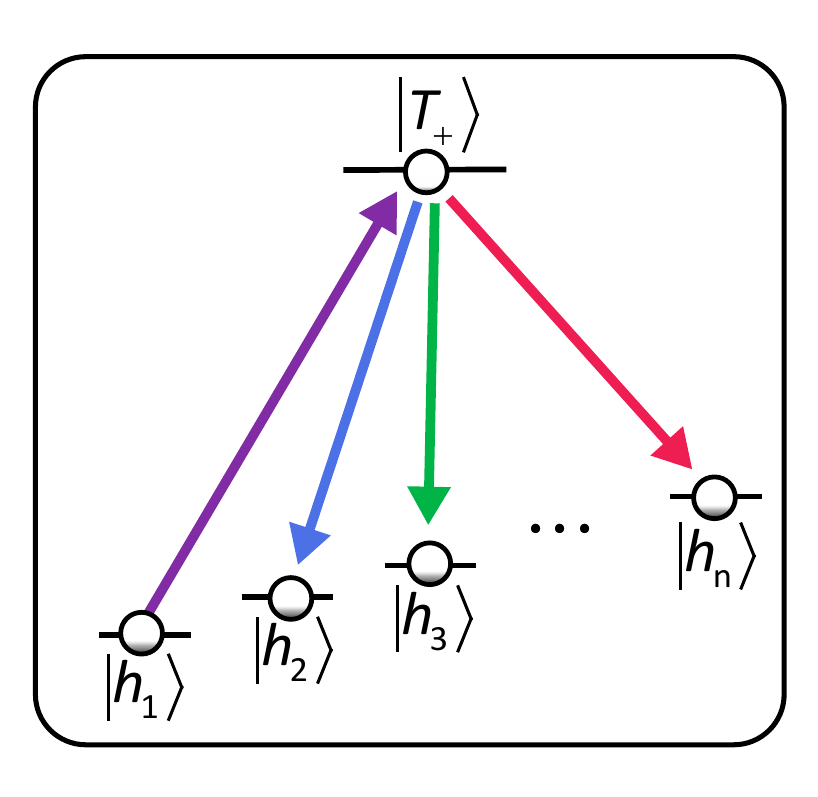}
\caption{The protocol to prepare superposition states composed of arbitrary hole orbital states.}
\label{Fig:superposition state}
\end{figure}

\newpage
\section{Filling time of the ground state and resolution of pump-probe technique}
\label{sec:filling time}

\begin{figure}[h]
\includegraphics[width=0.9\textwidth]{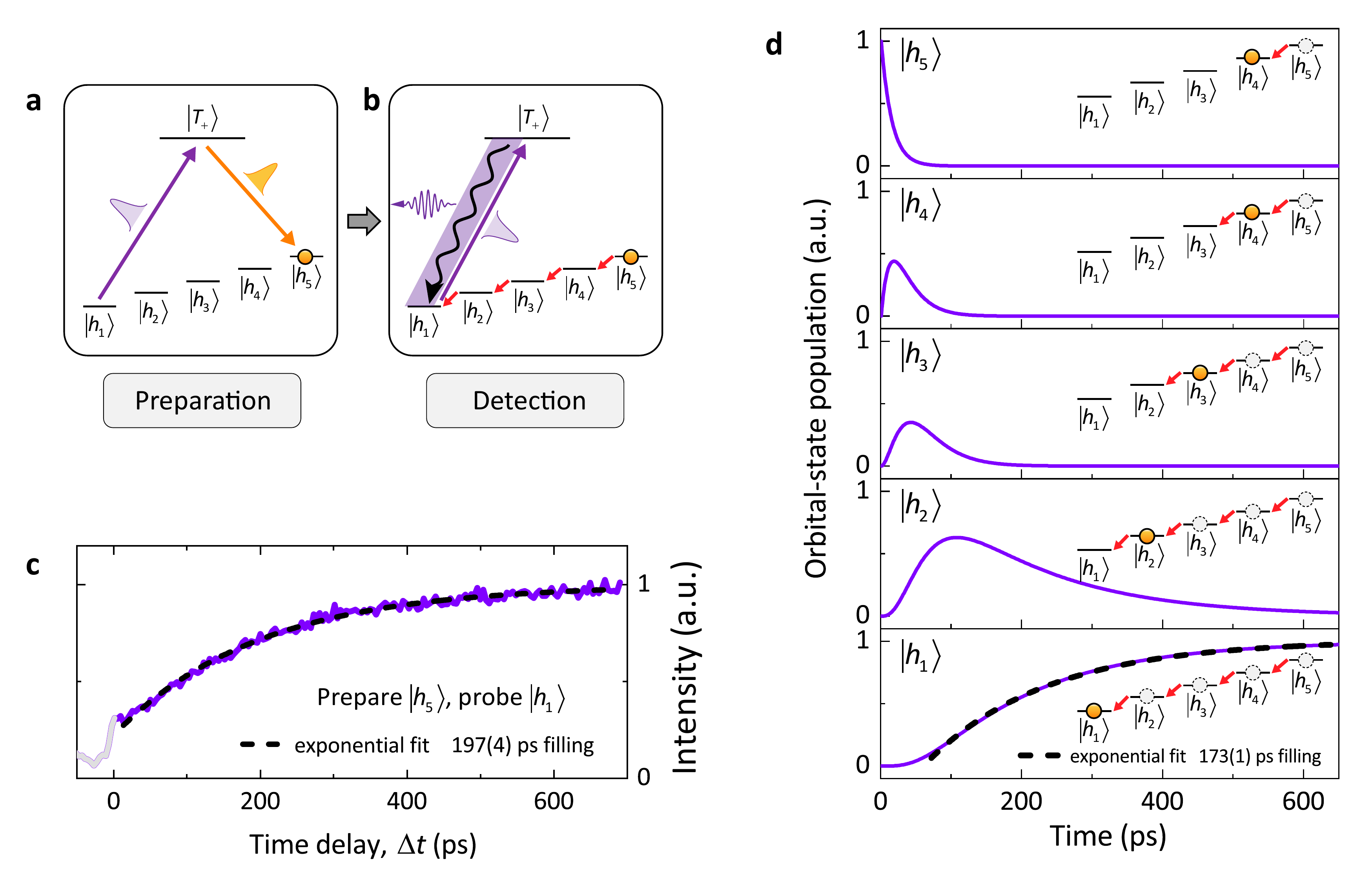}
\caption{\textbf{a}, Deterministic preparation of $\left|h_5\right>$. \textbf{b}, Probing the population filling of $\left|h_1\right>$. \textbf{c}, The population of $\left|h_1\right>$ as a function of the time delay between preparation and probing pulse. Black dashed line: single exponential fitting.  \textbf{d}, Simulated population evolution of orbital states from a cascade relaxation rate-equation model.}
\label{Fig:pre h5 probe h1}
\end{figure}

To obtain a deeper insight of the hole relaxation mechanism, we measure the filling dynamics of the hole-orbital ground state ($h_1$) after preparation of the $h_5$ orbital state. Supplementary Fig.~\ref{Fig:pre h5 probe h1}c shows the detected photon flux as a function of pulse interval $\Delta t$. A 197(4) ps filling time of $h_1$ is obtained from a single exponential fit. As shown in Supplementary Fig.~\ref{Fig:pre h5 probe h1}d, the filling curve can be well reproduced using a rate-equation model with experimentally determined $\tau_{h_2}$, $\tau_{h_3}$, $\tau_{h_4}$ and $\tau_{h_5}$, which yeilds a 173(1) ps filling time. These results indicate that the hole relaxation is dominated by a cascade decay process.

We note here that the temporal resolution of our pump-probe technique is fundamentally limited by the pulse duration of pump and probe pulses. The spectral width of pulses is $\sim$0.15 nm, corresponding to a transform-limited temporal duration of $T_p=$ $\sim$6 ps. Theoretically, convolving two 6-ps Gaussian pulses, we should get a duration of two-pulse overlap $T_o=\sqrt{2}T_p$.

Here, we utilize the AC Stark effect to characterize the temporal overlap of pump and probe pulses experimentally. As shwon in Supplementary Fig.~\ref{Fig:resolution}a and b, we use a $\pi$ pulse resonant with $\left|h_1\right>\leftrightarrow\left|T_+\right>$ transition to pump the QD, and then continuously vary the arrival time of another probe pulse (2.7 meV red-detuned). When the two pulses start to overlap, the AC Stark effect caused by the intense probe pulse shifts the QD-pump pulse resonance, thus reducing the efficiency of pumping. By varying the time delay of the two pulses and monitoring the intensity of $\left|T_+\right>\rightarrow\left|h_1\right>$ emission, we obtained an FWHM of two-pulse overlap $T_o=$ 9.5(1) ps very close to the theoretical expectation (8.5 ps), which corresponds to the temporal resolution of our pump-probe technique (Supplementary Fig.~\ref{Fig:resolution}c).

\newpage
\begin{figure}[h]
\includegraphics[width=0.7\textwidth]{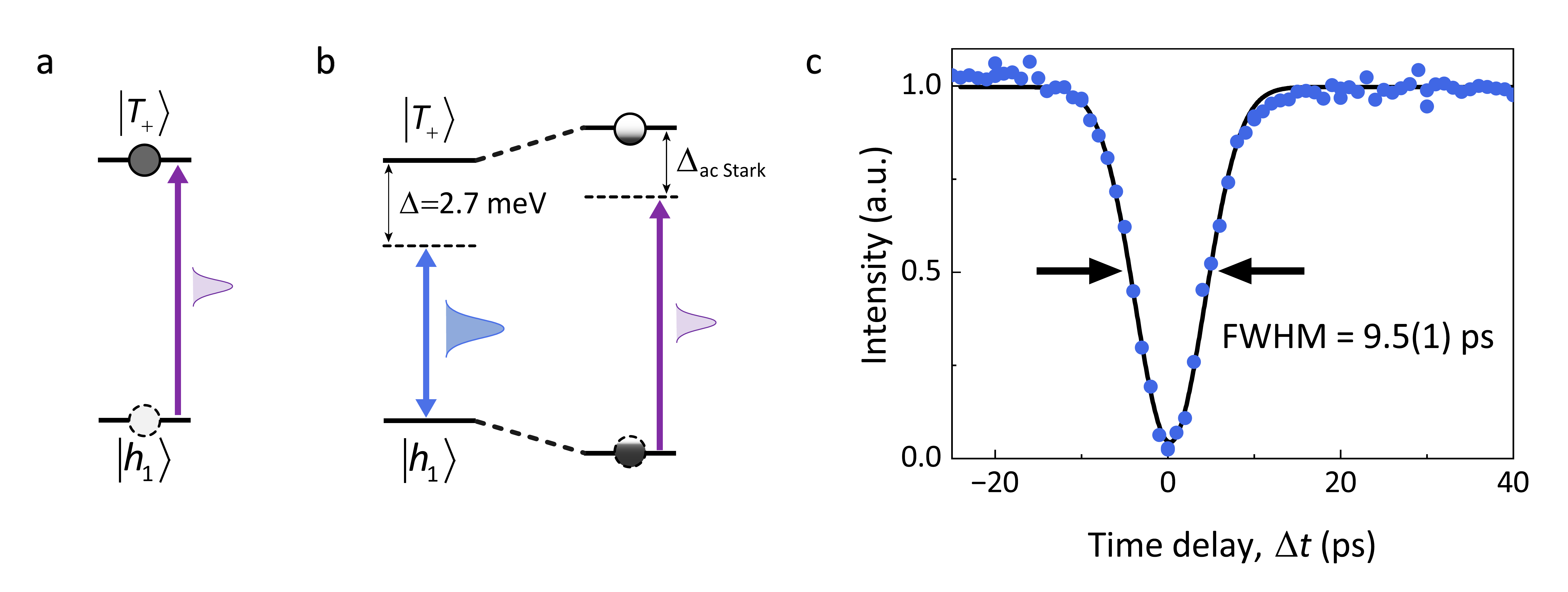}
\caption{The resolution of pump-probe technique. \textbf{a}, A $\pi$ pulse resonant with the $\left|h_1\right>\leftrightarrow\left|T_+\right>$ transition pumps the QD to $\left|T_+\right>$. \textbf{b}, The QD-pump pulse resonance is shifted after introducing another red-detuned pulse (blue) due to the AC Stark shift. \textbf{c}, The measured $\left|T_+\right>\rightarrow\left|h_1\right>$ photon flux as a function of time delay of two pulses, showing a Gaussian dip with an FWHM = 9.5(1) ps.}
\label{Fig:resolution}
\end{figure}

\newpage
\section{Modelling of the Ramsey interference}
\label{sec:Bloch equation}

In order to reproduce Ramsey fringes in Fig.~\ref{Ramsey}, we model the evolution of the two-level system (TLS) constructed by $\left|h_1\right> \textrm{and} \left|h_2\right>$. Although the $\left|\textit{T}_{\text{+}}\right>$ is involved in the actual experiment, since the interval between pulses 1 (3) and 2 (4) is constant (18 ps) and much smaller than the decoherence time of $\left|\textit{T}_{\text{+}}\right>$, $\left|\textit{T}_{\text{+}}\right>$ can be regarded as only a channel to detect the population of $\left|h_2\right>$ without affecting the system evolution. 

The initial Bloch vector after the pump $\pi/2$ pulse and first $\pi$ pulse $\textit{\textbf{M}}_{\textit{t}=0}$ reads:
\begin{equation}
    \textit{\textbf{M}}_{0}=
    \left[ {\begin{array}{cc}
    \textit{\textbf{M}}_{x} \\
    \textit{\textbf{M}}_{y} \\
    \textit{\textbf{M}}_{z}
    \end{array}}\right]
    _{0}=
    \left[ {\begin{array}{cc}
    1 \\
    0 \\
    0
    \end{array}}\right].
\label{initial_bloch}
\end{equation}
As shown in Supplementary Fig.~\ref{fig:bloch level}, after considering the longitudinal (population) relaxation, transverse (coherence) relaxation, and precession, the time-dependence Bloch vector leads to:
\begin{equation}
    \textit{\textbf{M}}_{\textit{t}}=
    \left[ {\begin{array}{cc}
    -e^{-\textit{t}(\frac{1}{2\tau_{\textit{h}_{\text{2}}}}+\frac{1}{\textit{T}_2^*})}\cos{2\pi\nu\textit{t}} \\
    -e^{-\textit{t}(\frac{1}{2\tau_{\textit{h}_{\text{2}}}}+\frac{1}{\textit{T}_2^*})}\sin{2\pi\nu\textit{t}} \\
    1-e^{\frac{-\textit{t}}{\tau_{\textit{h}_{\text{2}}}}}
    \end{array}}\right],
\label{evolutive_Bloch}
\end{equation}
where $\tau_{\textit{h}_{\text{2}}}$, $\textit{T}_2^*$, and $\nu$ is the lifetime of $\left|\textit{h}_{\text{2}}\right>$, and pure dephasing time of TLS and precession frequency, respectively. The third and fourth pulses further rotates the vector by $\pi/2$ around the y-axis, after which the Bloch vector leads to:
\begin{equation}
    \textit{\textbf{M}}_{\Delta\textit{t}}=
    \left[ {\begin{array}{cc}
    -e^{-\Delta\textit{t}(\frac{1}{2\tau_{\textit{h}_{\text{2}}}}+\frac{1}{\textit{T}_2^*})}\cos{2\pi\nu\Delta\textit{t}} \\
    1-e^{\frac{-\Delta\textit{t}}{\tau_{\textit{h}_{\text{2}}}}} \\
    -e^{-\Delta\textit{t}(\frac{1}{2\tau_{\textit{h}_{\text{2}}}}+\frac{1}{\textit{T}_2^*})}\sin{2\pi\nu\Delta\textit{t}}
    \end{array}}\right],
\label{final_Bloch}
\end{equation}
where $\Delta\textit{t}$ is the time delay between two pair of two-color pulse. Now, the final $\left|\textit{h}_{\text{2}}\right>$ state population is
\begin{equation}
    \textit{\textbf{P}}_{\left|\textit{h}_{\text{2}}\right>}=
    \frac{1}{2}(1-e^{-\Delta\textit{t}(\frac{1}{2\tau_{\textit{h}_{\text{2}}}}+\frac{1}{\textit{T}_2^*})}\cdot\sin{2\pi\nu\Delta\textit{t}}),
\label{population of h2}
\end{equation}
which is proportional to the detected photon flux. 
Substituting $1/T_2 = 1/(2\tau_{\textit{h}_{\text{2}}})+1/T_2^*$ and $T_2 = 276$ ps into equation \ref{population of h2}, we get the simulated envelope of the Ramsey fringes shown in Fig.~\ref{Ramsey}b.

\begin{figure}[h]
\includegraphics[width=0.28\textwidth]{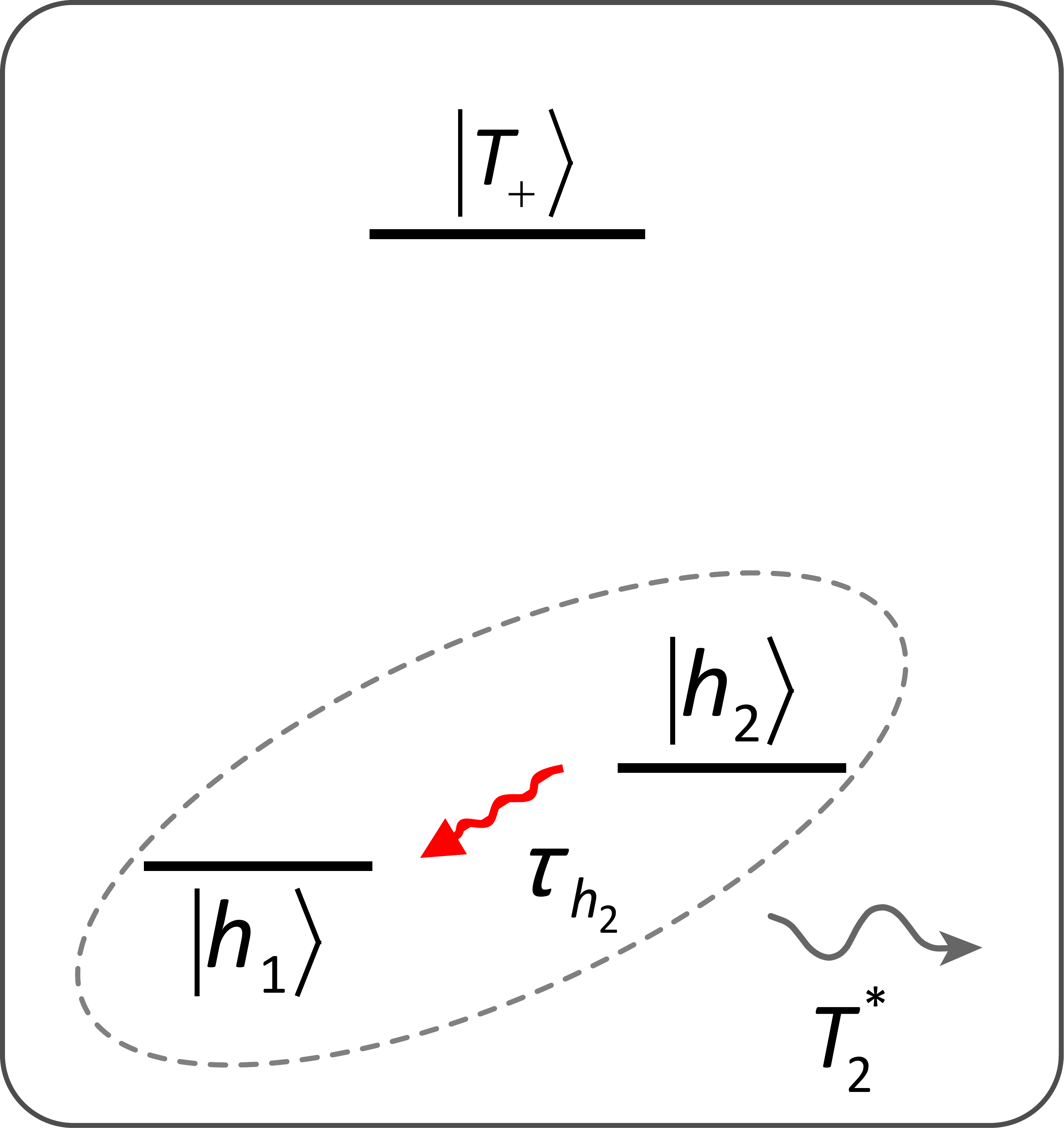}
\caption{The diagram of energy levels involved in the Ramsey interferometry.}
\label{fig:bloch level}
\end{figure}

\newpage
\section{Hole tunneling rate}
\label{tunneling}
To verify that the measured high-orbital state lifetime is related to the phonon-assisted relaxation rather than the hole tunneling process, we measure the bias dependence of hole lifetime and model the hole tunneling probability by the Wentzel-Kramers-Brillouin (WKB) method.

\subsection{Bias dependence of the high-orbital state lifetime}
We measure the lifetime of $\left|h_2\right>$ at different bias using the three-pulse pump-probe technique presented in Fig.~\ref{decay} in the main text. The experimental result is shown in Supplementary Fig.~\ref{fig:p- relaxation vs bias}. If the population decay of high-orbital hole is due to the hole discharging process, i.e., hole tunneling out from the QD, we would expect that the hole lifetime should have a strong dependence on the bias~\cite{Kurzmann2016a,Dalgarno2006,Korsch2020}. On the contrary, the measured $\tau_{h_2}$ stays around 148 ps in the entire bias range, suggesting the decay of $\left|h_2\right>$ population is not caused by the hole tunneling process.
\begin{figure}[h]
\includegraphics[width=0.45\textwidth]{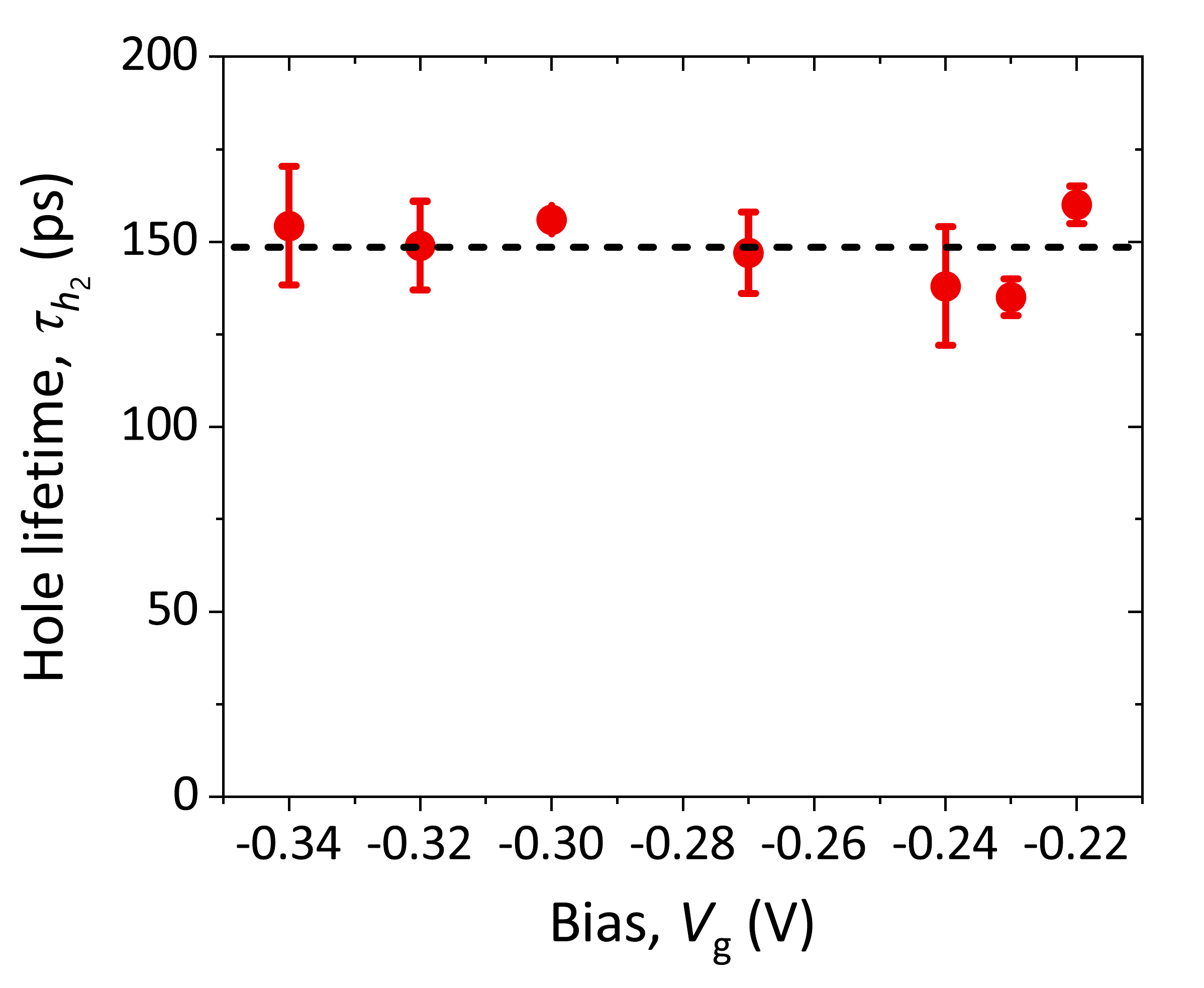}
\caption{$\left|h_2\right>$ lifetime as a function of bias $V_g$. The bias independence of the $\left|h_2\right>$ lifetime suggests that the population decay directly relates to the phonon-assisted relaxation instead of the hole tunneling process. dotted line: average hole lifetime (148 ps). Error bars arise from single-exponential fitting residual standard error.}
\label{fig:p- relaxation vs bias}
\end{figure}

\subsection{Calculation of hole tunneling rates}
\label{Sec:WKB}
We also calculate the hole tunneling rate based on the layer structure of the device. The calculated tunneling time of high-orbital holes is on the millisecond scale, much longer than the measured lifetime (161 ps or less) of the high-orbital hole states shown in the main text Fig.~\ref{decay}. Therefore, we conclude that the population decay of the high-orbital holes is mainly due to the phonon-assisted relaxation process.
The calculation uses the one-dimensional WKB approximation with a hole mass of $0.59~m_0$, and a temperature of 4 K. The hole transmission coefficient (also known as tunneling probability) through the $\rm Al_{0.33}Ga_{0.67}As$ barrier is given by \cite{Gehring}:
\begin{equation}
    TC(E) = \text{exp}(\frac{-2}{\hbar}\int_0^d\sqrt{2m_b(E-E_v(z))}dz),
\label{equ:WKB}
\end{equation}
where $\hbar$ is the reduced Planck constant, $d$ is the barrier width (see Supplementary Fig.~\ref{fig:wkb}b), $m_b$ is the hole effective mass in $\rm Al_{0.33}Ga_{0.67}As$ barrier~\cite{Levinshtein1996,Mar2011}, $E_v(z)$ is the position-dependent valence band edge of $\rm Al_{0.33}Ga_{0.67}As$ barrier, and $E$ is the energy of the hole orbital. Then, the hole tunneling time $\tau_t$ can be approximated by the following equation with the assumption that a hole with velocity $\nu$ collides back and forth in the QD at least 1/$TC$ times before tunneling out.
\begin{equation}
    \tau_t =\frac{1}{TC}\frac{2L_
    {QD}}{\nu},
\label{equ:tau}
\end{equation}
where $\nu$ is the hole velocity which can be calculated by $E$-$k$ relation ($\nu=(dE/dk)/\hbar$), and $L_{QD}$ is the height of the QD (about 5 nm). The simulated hole tunneling time at different orbitals are shown in Supplementary Fig.~\ref{fig:wkb}c.

\begin{figure}[h]
\includegraphics[width=1\textwidth]{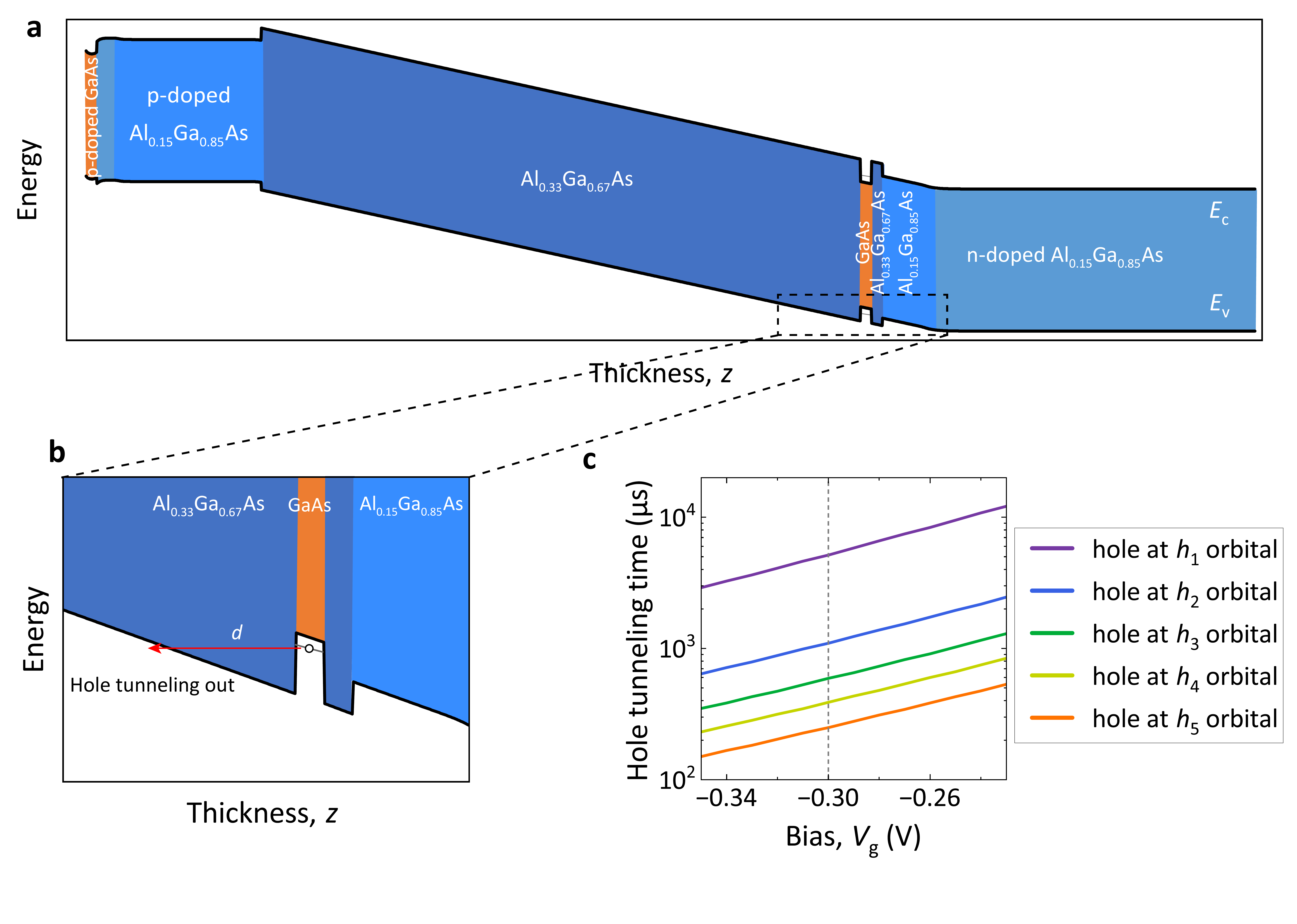}
\caption{WKB method simulation of hole tunneling time. \textbf{a}, Schematic band structure. $E_{\rm c}$: conduction band.  $E_{\rm v}$: valance band. \textbf{b}, Schematic diagram of the hole tunneling process, where the barrier width is denoted by $d$.  \textbf{c}, Hole tunneling time as a function of $V_g$ calculated by the WKB method for the hole at $h_1$, $h_2$, $h_3$, $h_4$, and $h_5$ orbitals. Vertical dotted line: bias $V_g=-0.3$ V, where we measured the hole lifetime.}

\label{fig:wkb}
\end{figure}

\putbib

\end{bibunit}
\end{document}